\documentclass[aps,prl,preprint,groupedaddress]{revtex4-1}
\usepackage{graphicx}% Include figure files
\usepackage{dcolumn}% Align table columns on decimal point
\usepackage{bm}% bold math

\begin{document}

% Use the \preprint command to place your local institutional report
% number in the upper righthand corner of the title page in preprint mode.
% Multiple \preprint commands are allowed.
% Use the 'preprintnumbers' class option to override journal defaults
% to display numbers if necessary
%\preprint{}

%Title of paper
\title{Parameter estimation
 of  default portfolios\\
 using the Merton model and Phase transition 
%\\
% using Beta binomial distribution
%\\
}

\author{Masato Hisakado}
\email{hisakadom@yahoo.co.jp} 
\affiliation{
 Nomura Holdings, Inc., Otemachi 2-2-2, Chiyoda-ku, Tokyo 100-8130, Japan} 

\author{Shintaro Mori}
\email{shintaro.mori@gmail.com}
\affiliation{
\dag Department of Mathematics and Physics,
Graduate School of Science and Technology, 
Hirosaki University \\
Bunkyo-cho 3, Hirosaki, Aomori 036-8561, Japan}

\date{\today}
\begin{abstract}
We discuss the parameter estimation of the probability of default (PD), the correlation between the obligors, and a phase transition.
  In our previous  work,
   we studied the problem using the beta-binomial distribution.
   A  non-equilibrium phase transition with an order parameter occurs
   when the temporal correlation decays by power law. 
   In this article, we adopt the Merton model, which uses an asset correlation  as
   the default correlation, and find that a phase transition  
   occurs 
   when the temporal correlation decays  by power law.
  When the power  index is less than one, the PD estimator
 converges slowly.
 Thus, it is difficult to estimate PD with limited historical data.
 Conversely, when the power index is greater than one, the convergence speed is inversely proportional to the number of samples.
We investigate the empirical default data history of several rating agencies.
The estimated  power index is in the slow convergence range when we use long history data.
This suggests that PD could have a long memory and 
that it is difficult to estimate parameters due to slow convergence.

\hspace{0cm}
\vspace{1cm}
%PACS numbers:  {02.50.Ga, 05.70.Fh, 89.65.Gh, 87.23.Kg} 
\end{abstract}

\maketitle
\bibliography{basename of .bib file}
%%%%%%%%%%%%%%%%%%%%%%%%%%%%%%%%%%%%%%%%%%%%%%%%
\newpage
\section{1. Introduction}

Anomalous diffusion is one of the most interesting topics
in sociophysics and econophysics \cite{galam,galam2,Man}.
The models describing such phenomena have a long memory \cite{Bro,W2,G,M,hod,hui,sch} and
show several types of phase transitions.
In our previous work, we investigated voting models for an information cascade 
\cite{Mori,Hisakado2,Hisakado3,Hisakado35,Hisakado4,Hisakado5,Hisakado6}.
This model has two types of phase transitions.
One is the information cascade transition, which is similar to the
 phase transition of the Ising model \cite{Hisakado3}
 that shows whether a distribution converges.
The other phase transition is the convergence transition of the
super-normal diffusion that corresponds to an anomalous diffusion  \cite{Hod,Hisakado2}.

In financial engineering, several products have been invented to hedge risks.
The credit default swap (CDS) is one tool used to hedge credit risks and is a single name credit derivative that targets the default of one single obligor.
Synthetic collateralized debt obligations (CDOs) are financial innovations that securitize portfolios of assets, which, in the 2000s, became the trigger of the great recession in 2008.
These products provide protections against a subset of the total loss on a credit portfolio in exchange for payments.
They provide valuable insights into market implications on default dependencies and the clustering of defaults. This final aspect is important because the difficulties in managing credit events depend on  correlations.

Estimations of the probability of default (PD) and correlation between the obligors have been obtained from empirical studies on historical data. These two parameters are important for pricing financial
products such as synthetic CDOs \cite{M2010,M2008,Sch}. Moreover,
they are important to financial institutions for portfolio management and are
 called "long-run PDs" in the regulations. When defaults are minimal,
it is not easy to estimate these parameters when there is a correlation \cite{Tas,FSA}.

In this work, we study a Bayesian estimation method using the
Merton model. Under normal circumstances, the Merton
model incorporates default correlation by 
the correlation of asset price movements (asset correlation), which
is used to estimate the PD and the correlation.
A Monte Carlo simulation is an appropriate tool to estimate the
parameters, except under the 
limit of large homogeneous portfolios 
\cite{Sch}. 
In this case, the distribution becomes a Vasciek distribution
that can be calculated analytically \cite{V}. 

In our previous paper,
we discussed  parameter estimation using the beta-binomial distribution　with default correlation and considered a multi-year case with a temporal correlation \cite{Hisakado6}.
A non-equilibrium phase transition, like that of the Ising model, occurs when the temporal correlation decays
by power law. 
In this study, we discuss  a phase transition when we use the Merton model.  
When the
power index is less than one,
the estimator distribution of the PD
 converges slowly to the delta function.
Alternatively, when the power index is greater than one, the
convergence is the same as that of the normal case.
When the distribution slowly converges, it takes time to estimate the 
PD with limited data.

To confirm the decay form of the temporal correlation,
we investigate empirical default data.
We confirm the estimation of the power index in the slow convergence range.
This demonstrates that even if there exists adequate historical data,
it will take time to correctly estimate the parameters of PD, asset correlation, and temporal correlation.

%%%%%%%%%%%%%%%%%%%%%%%%%%%%%%%%%

The remainder of this paper is organized as follows.
In Section 2, we introduce the stochastic process of the Merton model 
 and consider the convergence of the PD estimator. 
In Section 3, we apply Bayesian estimation approach to the empirical
data of default history using the Merton model and confirm its parameters. 
The  estimated parameter is in the slow convergence phase.
Finally, the conclusions are presented in Section 4.

\section{2. Asset correlation and default correlation}

In this section we consider whether the time series of a stochastic process
using the Merton model converges \cite{Mer}. We show that 
the convergence is intimately related
to the phase transition.
Using this conclusion, we discuss if we can estimate the
parameters.

Normal random variables, $S_t$, are hidden variables that explain
the status of the economics and $S_t$ affects all obligors in the $t$-th year.
In order to introduce the temporal correlation of the defaults
from different years, let $\{S_t,1\le t \le T\}$ be the time series of the stochastic
  variables of the correlated normal distribution with the following
 correlation matrix $\Sigma$:
\begin{equation}
\Sigma\equiv\left(
    \begin{array}{ccccc}
     1 &  d_1 & d_2 &\cdots& d_{T-1} \\
   d_1      &  1 &  d & \ddots&\vdots \\
  \ddots & \ddots & \ddots  & \ddots&\ddots  \\
 \vdots & \ddots & \ddots  & \ddots &d\\
    d_{T-1}&  \cdots &  d_2&  d  & 1  \\
    \end{array}
  \right),
\end{equation}
 where $(S_1,\cdots,S_T)^T\sim \mbox{N}_{T}(0,\Sigma)$. 
In this work, we consider two cases of temporal correlation: exponential  decay, $d_{i}=\theta^i,0\le
\theta\le 1$,
and power decay, $d_{i}=1/(i+1)^{-\gamma},\gamma\ge 0$.
The exponential decay corresponds to short memory and the power decay corresponds to
intermediate and long memories \cite{Long}.
Without loss of generality, we assume the number of obligors   in the $t$-th year
is constant and we denote it as $n$.

The asset correlation, $\rho_A$, is the parameter that describes the 
correlation between the value of the assets of the obligors in the same year.
We consider the $i$-th asset value, $\hat{U}_{it}$, at time  $t$, to be
\begin{equation}
    \hat{U}_{it}=\sqrt{\rho_A}S_t+\sqrt{1-\rho_A}\epsilon_{it},
\end{equation}
where $\epsilon_{it}\sim \mbox{N}(0,1)$ is i.i.d.
By this formulation, the equal-time correlation of $U_{it}$ is $\rho_{A}$.
The discrete dynamics of the process is described by
\begin{equation}
    X_{it}=1_{\hat{U}_{it} \leq Y},
\label{prpcess}
\end{equation}
where $Y$ is the threshold  and $1 \leq i \leq n$.
When $X_{it}=1 (0)$, the $i$-th obligor in the $t$-th year 
is default (non-default).
Eq.(\ref{prpcess}) corresponds to the conditional default probability for $S_{t}=S$
as
\begin{equation}
G(S)\equiv \mbox{P}(X_{it}=1|S_{t}=S)=\Phi \left(\frac{Y-\sqrt{\rho_A} S}{\sqrt{1-\rho_A}}\right),
 \label{V}
\end{equation}
where $\Phi(x)$ is the standard normal distribution, 
$G(S_t)$ is the distribution of the default probability
during the $t$-th year in the portfolio,
and the average PD is $p=\Phi(Y)$, which corresponds to ``long run PDs".

The default correlation, $\rho_{D}$, is 
\[
\rho_{D}=f(\rho_{A})\equiv \frac{\mbox{P}(X_{it}=1 \cap X_{jt}=1)-p^2}{p(1-p)}
=\frac{\Phi_{2}((\Phi^{-1}(p),\Phi^{-1}(p)),\rho_{A})-p^2}{p(1-p)},
\]
where $\Phi_{2}$ denotes the bivariate normal distribution
with standardized marginals.
We define the mapping function, $\rho_D=f(\rho_A)$,
between the default correlation $\rho_D$ and the asset correlation $\rho_A$.
Note that the mapping function, $\rho_D=f(\rho_A)$, depends on $p$.
By the temporal correlation of $S_t$, we have
the asset correlation of the asset values at
different times as
\begin{equation}
\mbox{Cor}(U_{it},U_{jt'})=\rho_A \Sigma_{t,t'},
\label{matrix}
\end{equation}
where $\rho_A$ is the asset correlation and $\Sigma$
is the correlation matrix for $S_{t}$, $t=1,\cdots,T$.
In Appendix A we explain how to calculate Eq.(\ref{matrix}).
The default correlation
between $X_{it}$ and $X_{jt'}$ is given by 
\[
\frac{\mbox{P}(X_{it}=1 \cap X_{jt'}=1)-p^2}{p(1-p)}
=\frac{\Phi_{2}((\Phi^{-1}(p),\Phi^{-1}(p)),\rho_{A}d_{|t-t'|})-p^2}{p(1-p)}
=f(\rho_{A}d_{|t-t'|}).
\]

We are interested in the unbiased estimator of PD, $Z(T)\equiv \sum_{t,i}X_{it}/(nT)$, and
the limit $\lim_{T\to \infty}Z(T)$.
As the covariance of $X_{it}$ and $X_{jt'}$ is $p(1-p)f(\rho_{A}d_{|t-t'|})$,
the variance of $Z(T)$ is
\[
\mbox{V}(Z(T))=p(1-p)\frac{1}{nT}+p(1-p)\frac{(n-1)}{nT}f(\rho_{A})
+2p(1-p)\frac{1}{T^2}\sum_{i=1}^{T-1}f(\rho_{A}d_i)(T-i).
\]
The first term is from the binomial distribution, the second term is from the default correlation
in the same year, and  the third term is from the temporal correlation.
In the limit $T\to \infty$, the first two terms disappear and the convergence of
the estimator $Z(T)$ is governed by the third term.

We study the asymptotic behavior of $f(\rho_{A}d_{i})$ for large $i$. $f(\rho_A)$ is explicitly given as
\begin{eqnarray}
f(\rho_A)&=&\frac{\Phi_{2}((\Phi^{-1}(p),\Phi^{-1}(p)),\rho_A)-p^2}{p(1-p)} \nonumber \\
&=&\frac{1}{p(1-p)}\left(
\frac{1}{2\pi (1-\rho_A^2)}\int_{-\infty}^{\Phi^{-1}(p)} dx \int_{-\infty}^{\Phi^{-1}(p)} dy
\exp(-\frac{1}{2(1-\rho^2)}(x^2+y^2-2\rho_A xy))-p^2\right). \nonumber 
\end{eqnarray}
As we assume that $d_i$ decays to zero for large $i$, we expand $f(\rho_A)$ at $\rho_A=0$
as
\[
f(\rho_A)=\frac{\rho_A}{2\pi p(1-p)}(\int_{-\infty}^{\Phi^{-1}(p)}\exp(-\frac{1}{2}x^2)dx)^2+O(\rho_A^2).
\]
We denote the coefficient as $A$ and $f(\rho_A)\simeq A\rho_A$. $A$ is defined as
\begin{equation}
A\equiv \frac{1}{2\pi p(1-p)}\left(\int_{-\infty}^{\Phi^{-1}(p)}\exp(-\frac{1}{2}x^2)dx\right)^2>0 \label{eq:A}.
\end{equation}
Hence, we can confirm $A>0$.
For large $i$ and $d_i\to 0$, we have the asymptotic behavior of the default correlation as
\[
C(t)\equiv \mbox{Cor}(X_{is},X_{j(s+t)})=f(\rho_{A}d_t)\simeq A\rho_{A}d_t.
\]
We note that the default correlation $f(\rho_{A}d_i)$ obeys the same decay law as that of  $d_i$ for large $i$ and $d_i\to 0$.

\begin{figure}[h]
\begin{center}
\begin{tabular}{ccc}
\includegraphics[clip, width=5cm]{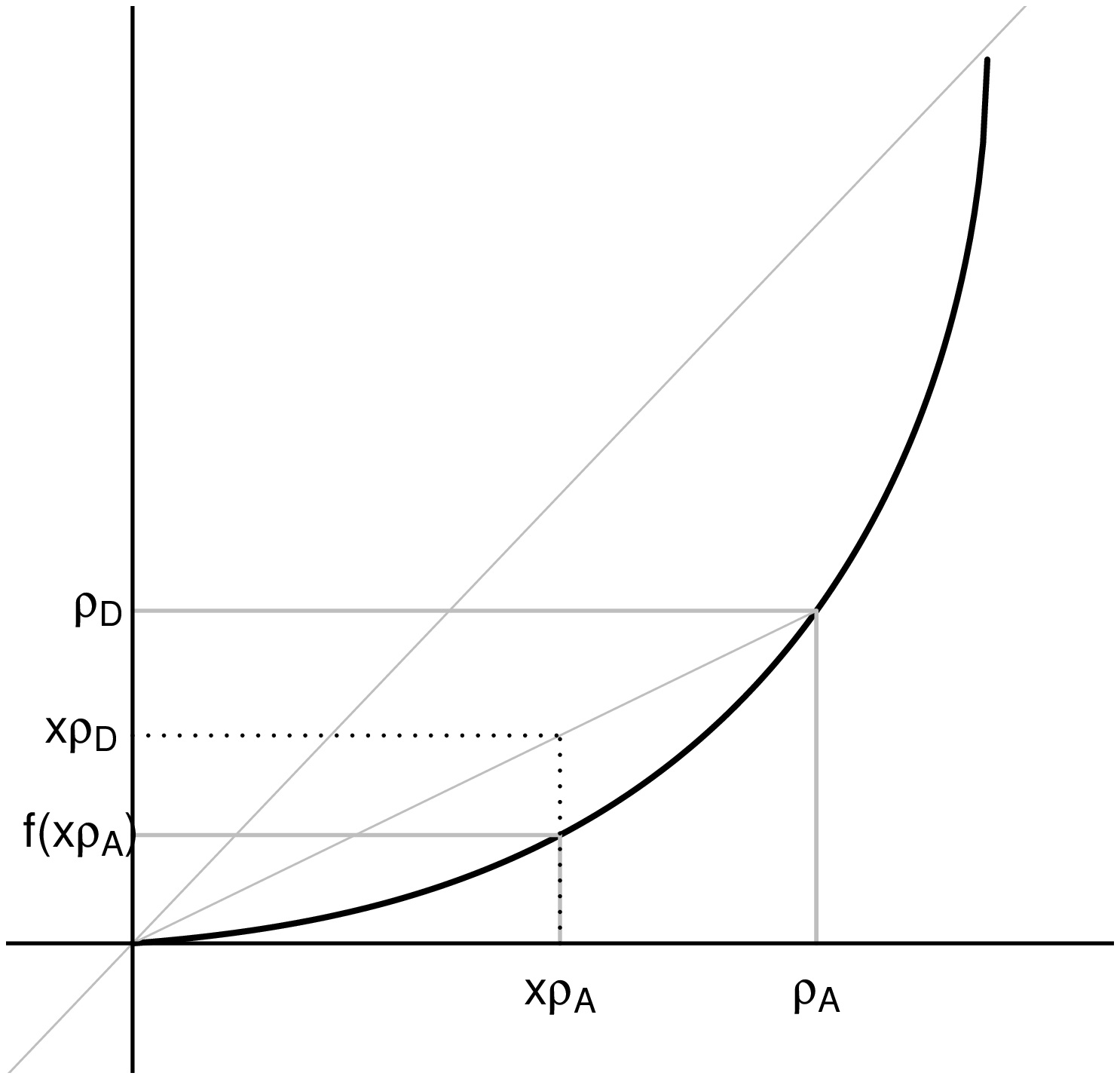}
\includegraphics[clip, width=5cm]{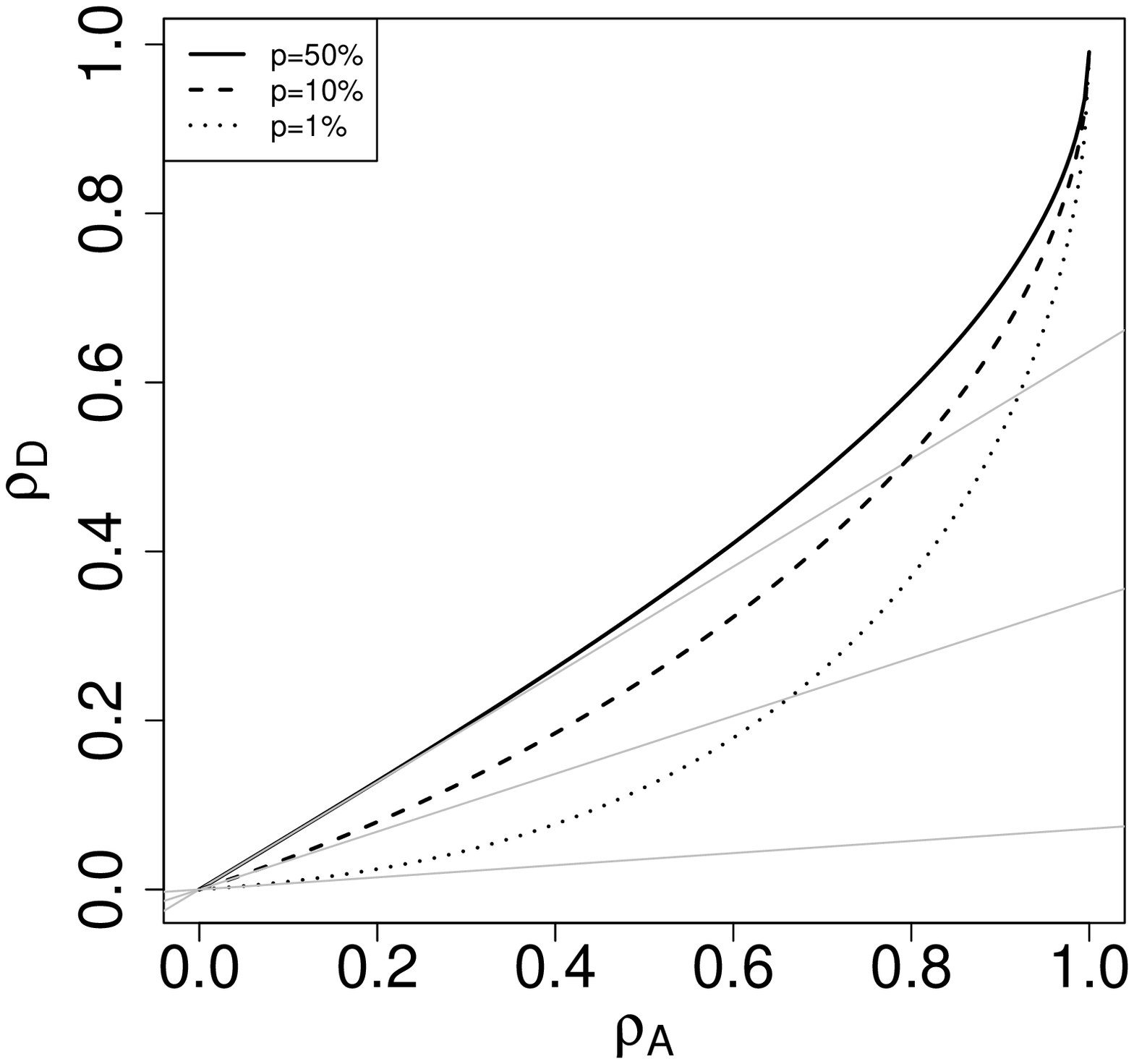}
\includegraphics[clip, width=5cm]{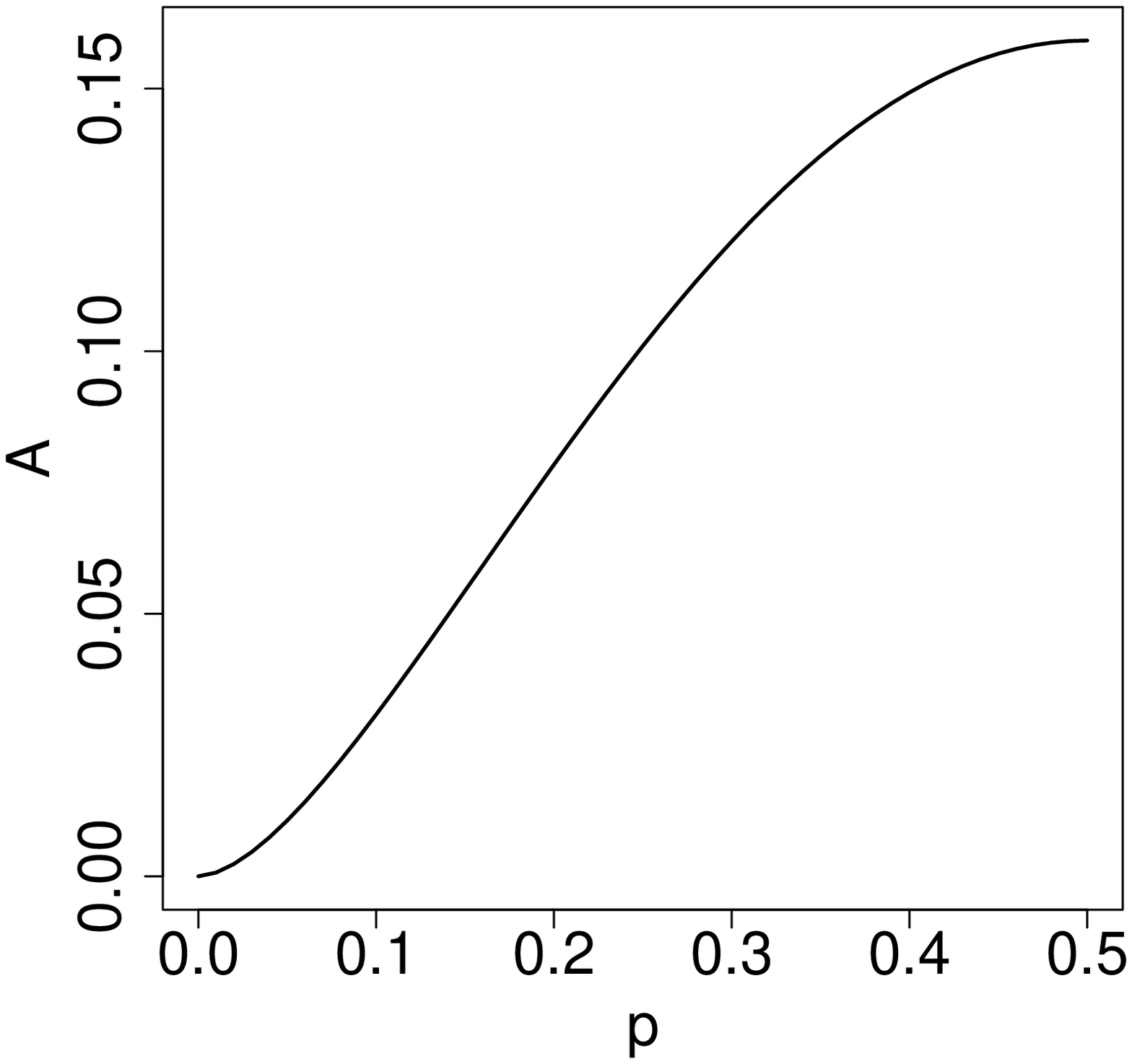}
\end{tabular}
\caption{(a) Plot of  the
  mapping function $f(\rho_{A})$ vs. $\rho_{A}$,
  which shows the relation between the
  asset correlation $\rho_A$ and the default correlation $\rho_D$.
  $f(\rho_{A})$ is downward convex and 
  $f(x\rho_{A})\le x f(\rho_{A})$ for $0\le x\le 1$.
  (b) We set $p=0.5$(solid), $0.1$(broken) and $0.01$(dotted).
  The gray lines show the lines with a slope of $A$ in Eq.(\ref{eq:A}).
  (c) Plot of $A$ vs. $p$. }
%\label{p_vs_Tau_r_q06}
\label{rho}
\end{center}
\end{figure}

We plot the mapping function $\rho_{D}=f(\rho_{A})$ in the $(\rho_{A},\rho_{D})$-plane in Fig. \ref{rho} under the
conditions $f(0)=0$ and $f(1)=1$.
The straight lines $\rho_D=A\rho_A$ with $A$ from  Eq.(\ref{eq:A})  are plotted  in Fig. \ref{rho} (b). 
 We confirm that 
$A$ in Eq.(\ref{eq:A}) is the same as the slope of the tangent line at point $(\rho_A,\rho_D)=(0,0)$.
We plot the relation between $p$ and $A$ in Fig. \ref{rho}(c).
From the convexity of $f(\rho_{A})$ in Fig. \ref{rho}(a),
we find the following inequality:
\[
A\rho_{A}d_i< f(\rho_{A}d_i)< f(\rho_{A})d_i=\rho_{D}d_i \,\,,\,\, \mbox{for}\,\,\,\,\rho_{A},\rho_{D} >0,0<d_i<1
\]

Using this inequality, we form the upper bound for $V(Z(T))$ as:
\begin{eqnarray}
\mbox{V}(Z(T))&=&p(1-p)\frac{1}{nT}+p(1-p)\frac{(n-1)}{nT}f(\rho_{A})
+2p(1-p)\frac{1}{T^2}\sum_{i=1}^{T-1}f(\rho_{A}d_i)(T-i) \nonumber \\
& \leq &
p(1-p)\frac{1}{nT}+p(1-p)\frac{(n-1)}{nT}f(\rho_{A})
+2p(1-p)\rho_{D}\frac{1}{T^2}\sum_{i=1}^{T-1}d_i (T-i). \nonumber
\end{eqnarray}
The lower bound is then
\[
V(Z(T))
\geq
p(1-p)\frac{1}{nT}+p(1-p)\frac{(n-1)}{nT}f(\rho_{A})
+2p(1-p)A\rho_{A}\frac{1}{T^2}\sum_{i=1}^{T-1}d_i(T-i).
\]
In both the upper and lower bounds, their third term is proportional to $\frac{1}{T^2}\sum_{i=1}^{T-1}d_i(T-i)$.
Thus, we can estimate the asymptotic behavior of $V(Z(T))$ by
the following expression:
\begin{equation}
\mbox{V}(Z(T))= p(1-p)\frac{1}{nT}+p(1-p)\frac{(n-1)}{nT}\rho_{D}
+2p(1-p)c\frac{1}{T^2}\sum_{i=1}^{T-1}d_i(T-i),
\end{equation}
where $c$ is a positive constant and $\rho_D>c>A\rho_A$.

\subsection{2.1 Exponential temporal correlation}
In this subsection, we study the convergence of $Z(T)$
for the exponential decay model $d_i=\theta^i,\theta\leq 1$:
\begin{equation}
V(Z(T))\simeq 
p(1-p)\frac{1}{nT}+p(1-p)\frac{(n-1)}{nT}f(\rho_{A})
+2p(1-p)c\frac{1}{T^2}\sum_{i=1}^{T-1}\theta^i (T-i). \nonumber
\end{equation}
The first two terms on the right-hand side (RHS) behave as $\propto 1/T$ and, thus, 
converge to 0 in the limit $T\to \infty$.
In the case that $\theta \neq 1$, the third term is
\[
2p(1-p)c\frac{1}{T^2}[T\frac{1-\theta^{T-1}}{1-\theta}
+\frac{(T-1)\theta^{T-1}(1-\theta)-(1-\theta^{T-1})\theta}{(1-\theta)^2}  
\propto 1/T
\]
and it converges to 0 in the limit $T\to \infty$.
In addition, $C(t)\simeq A\rho_{A}\theta^{t}$ for large $t$.
We conclude that as the number of data samples increases, the distribution of $Z(T)$
converges to a delta function and therefore,  
PD can be estimated empirically.

\begin{figure}[htbp]
\begin{tabular}{ccc}    
(a)\includegraphics[width=4.5cm]{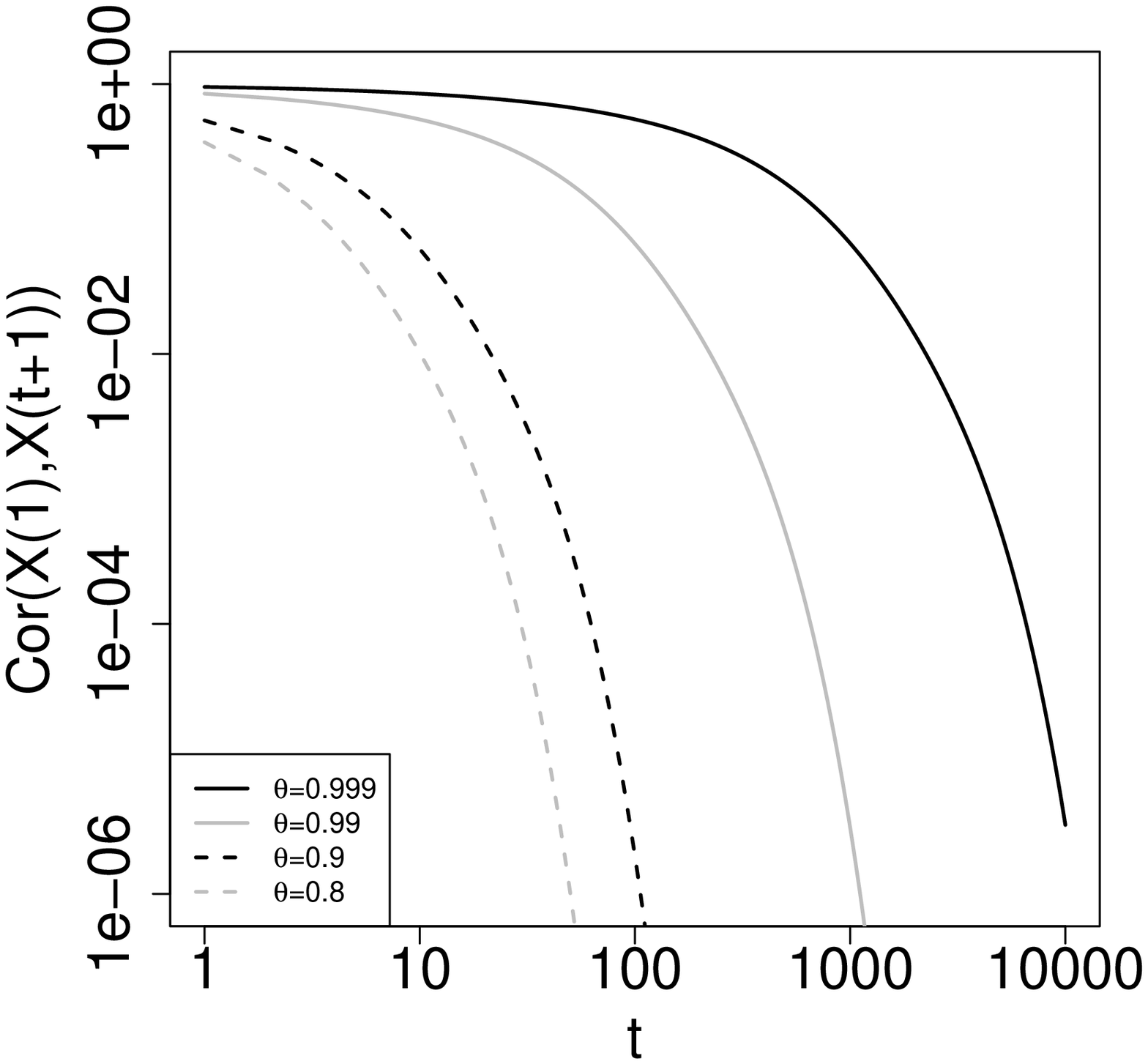} &
(b)\includegraphics[width=4.5cm]{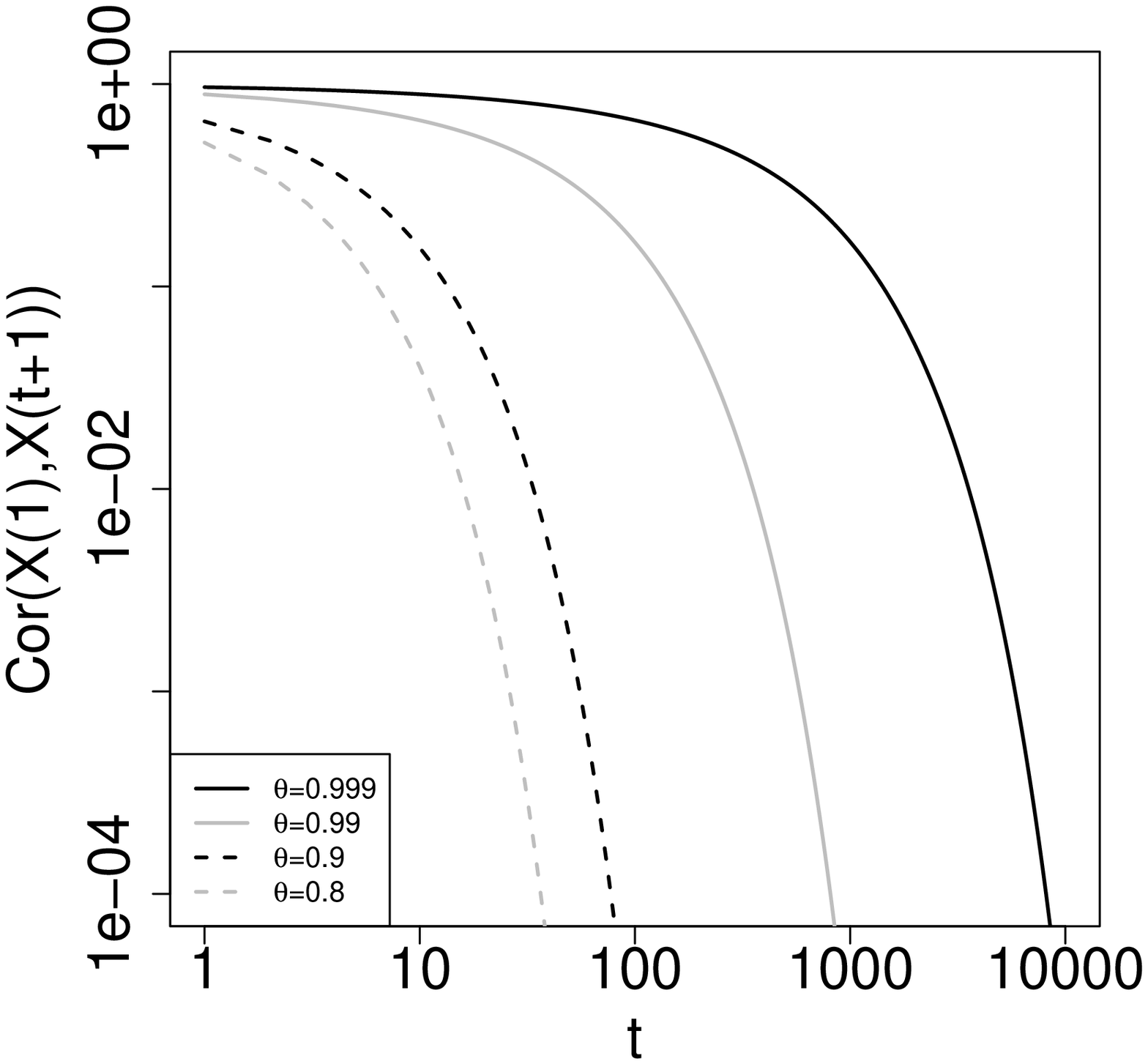} &
(c)\includegraphics[width=4.5cm]{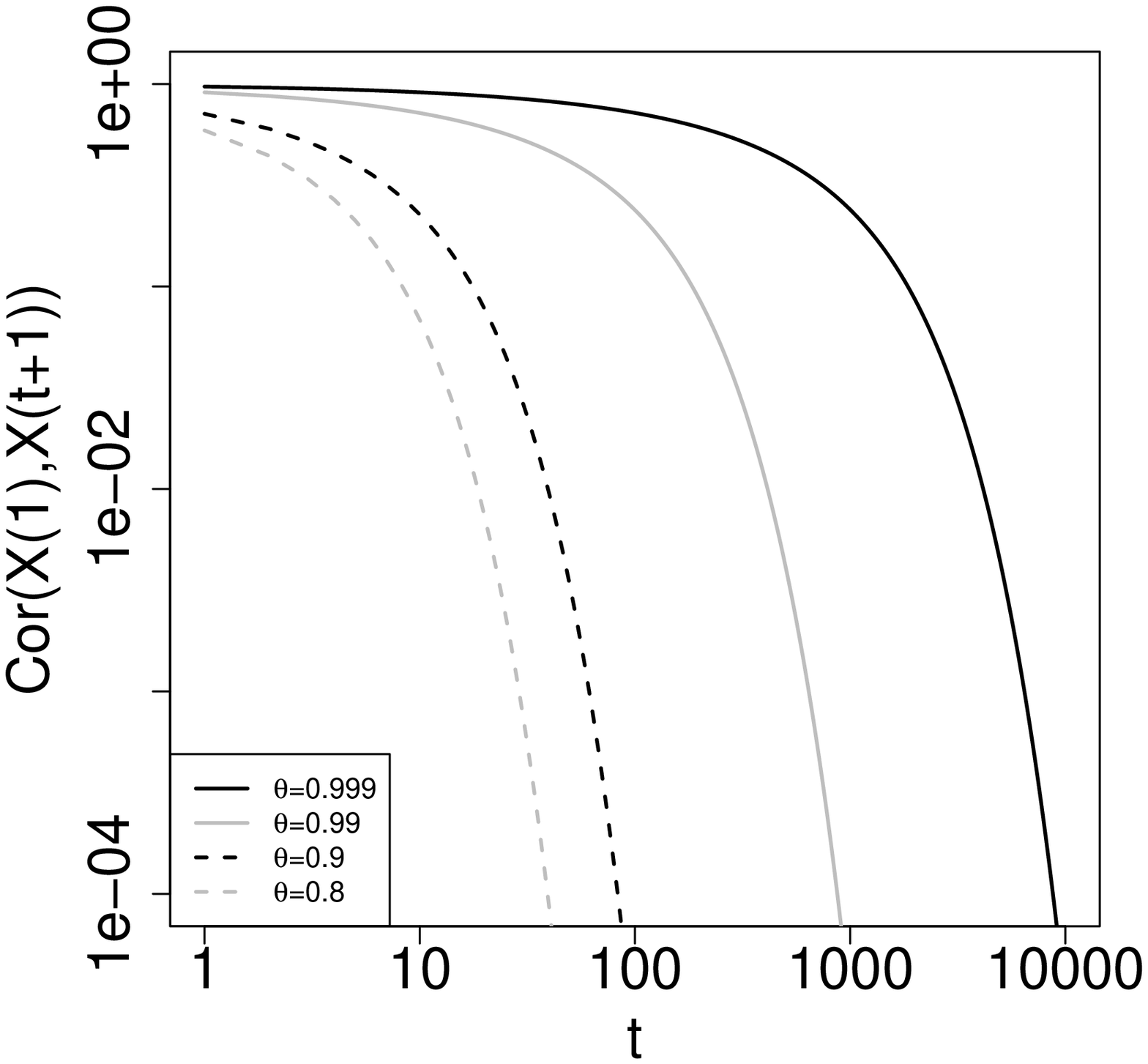} \\
(d)\includegraphics[width=4.5cm]{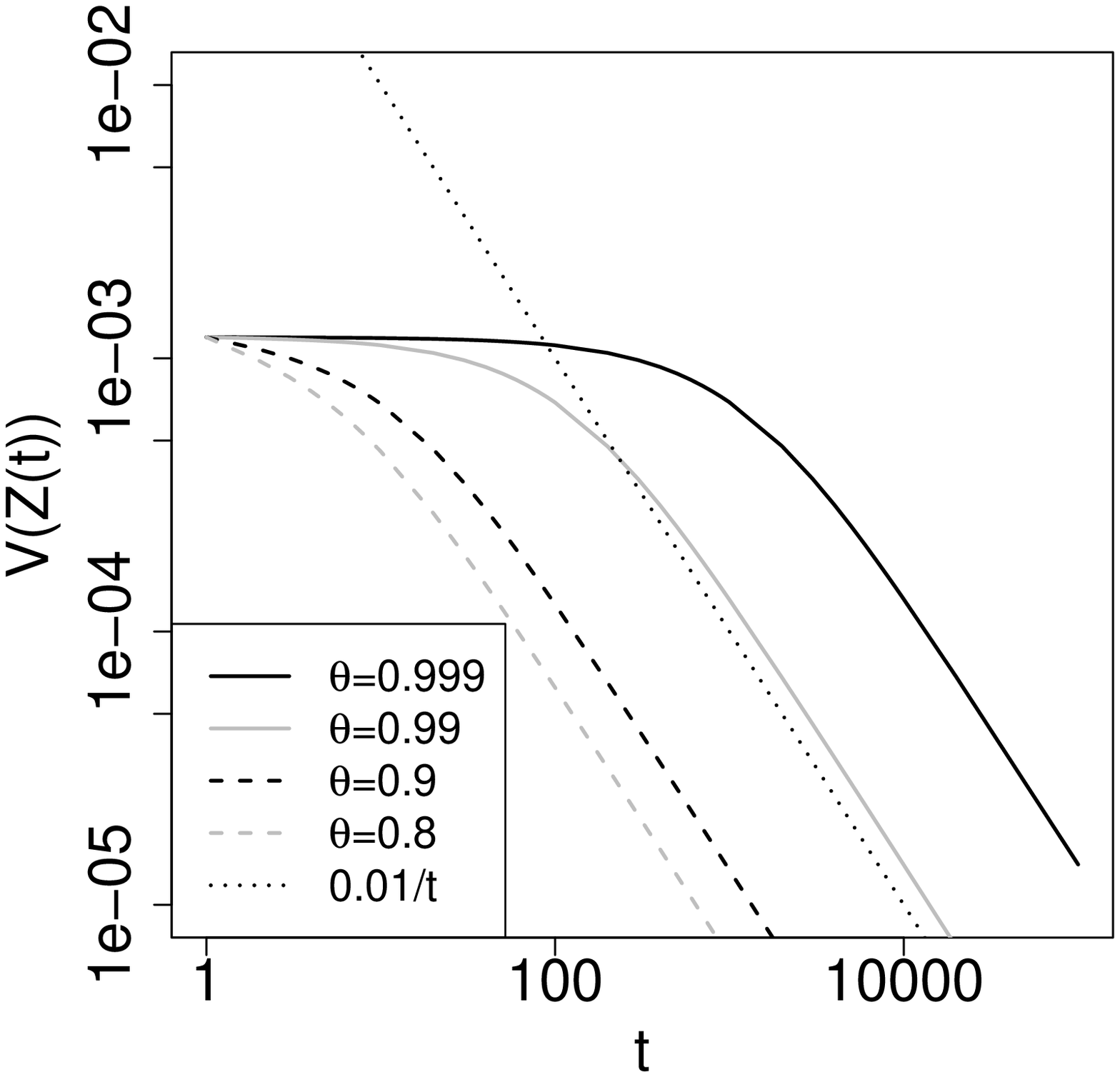} &
(e)\includegraphics[width=4.5cm]{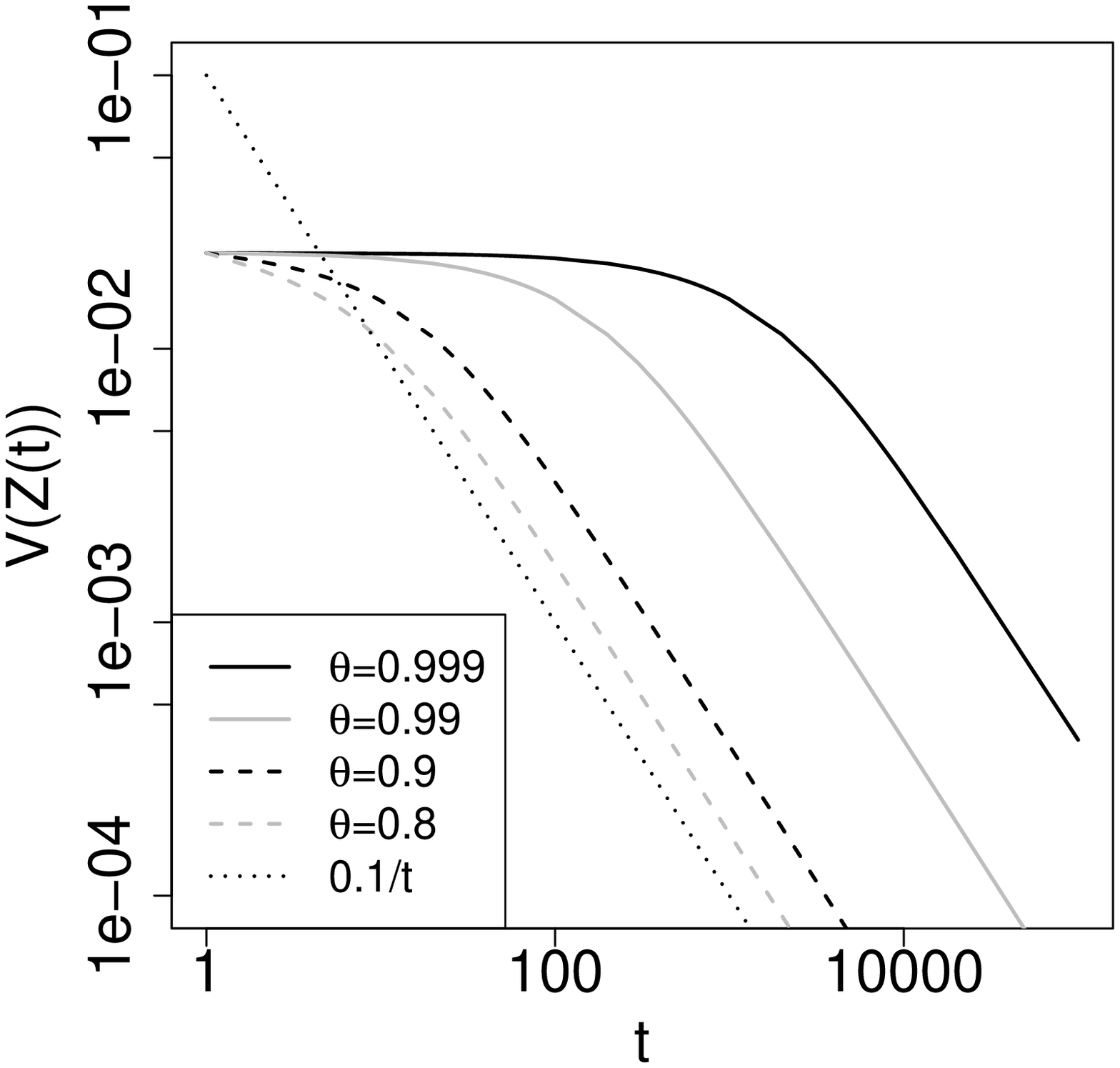} &
(f)\includegraphics[width=4.5cm]{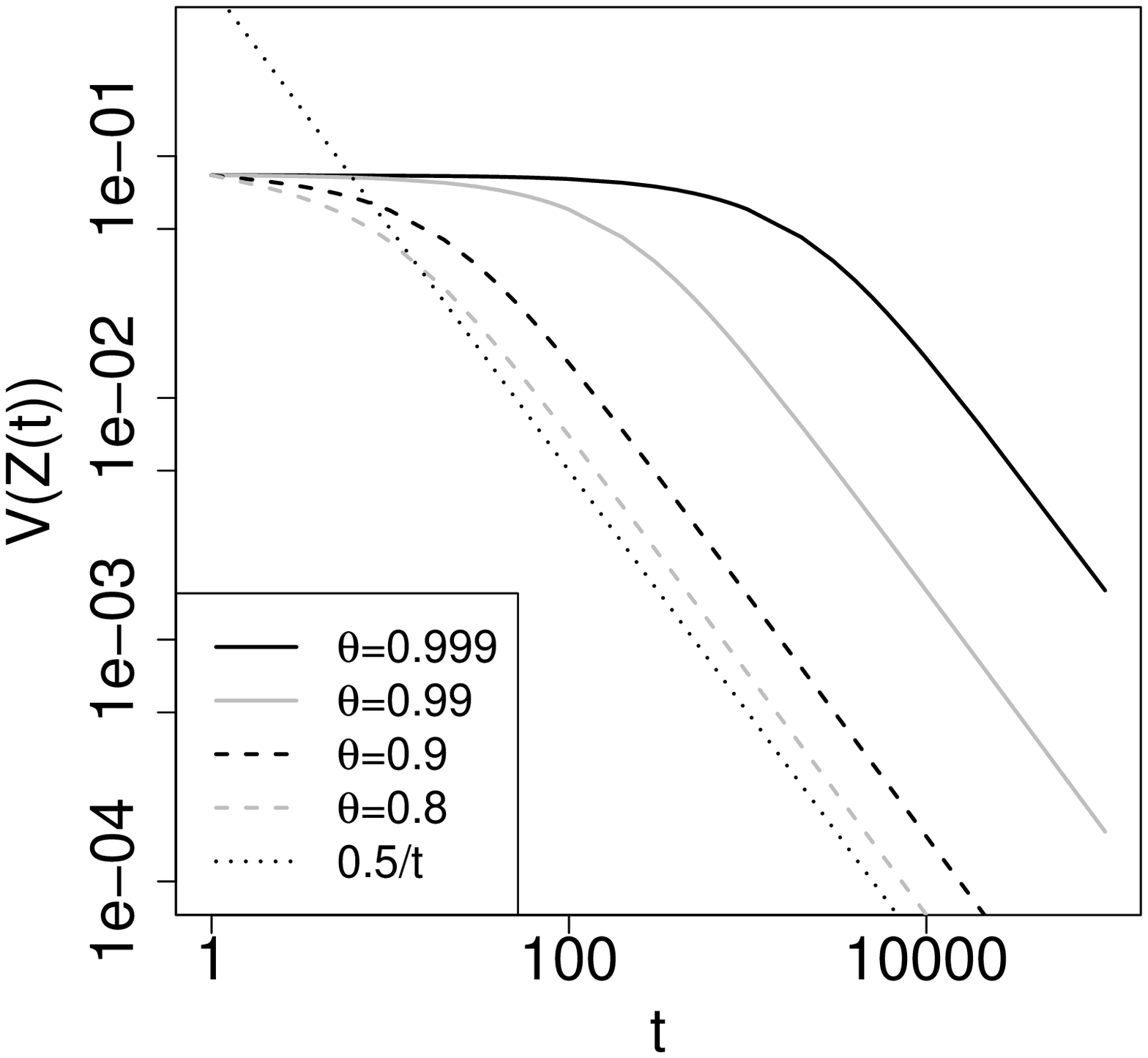} \\
\end{tabular}  
\caption{Plots of (a),(b),(c) $C(t)$ and (d), (e), (f) $V(Z(t))$
  vs. $t$, for $\theta \in \{0.8,0.9,0.99,0.999\}$, Exponential decay case.
  The PDs are  $1\%$ for (a) and (d), $10\%$ for (b) and (e), and $50\%$ for (c) and (f).
 }
\label{EXP}
\end{figure}

Thus, we calculate $C(t))=f(\rho_{A}\theta^{t})$ and 
$V(Z(t))$ numerically for $t\le 10^5$.
We set $\rho_{A}=0.5,n=10^4,p=0.5,0.1,0.01$ and $\theta=0.8,0.9,0.99,0.999$.
Figure \ref{EXP} (a)-(c) shows the double logarithmic plot of $C(t)$ vs. $t$.
Here, it is clearly seen that $C(t)$ decays exponentially.
Figure \ref{EXP} (d)-(f) shows the plot of V$(Z(t))$ vs. $t$.
For all $\theta<1\in \{0.8,0.9,0.99,0.999\}$, $V(Z(t))$ decays as $1/t$. 
When $\theta=1$, there is no temporal correlation decay case and all obligors are correlated $\rho_A$. Hence, there is no phase transition for $\theta<1$.

\subsection{2.2 Power temporal correlation}

In this subsection, we consider
the power decay case $d_i=1/(i+1)^{\gamma},i=1,2,\cdots$,
where $\gamma\ge 0$ is the power index.
The power correlation affects the number of defaults for long periods of time.
The ranges $\gamma\leq1$  and $\gamma>1$ are called long memory and intermediate memory, respectively \cite{Long}.
On the other hand, the exponential decay affects short periods of time and it is called short memory.
The asymptotic behavior of $V(Z(T))$ is given as:
\[
\mbox{V}(Z(T))\simeq p(1-p)\frac{1}{nT}+p(1-p)\frac{(n-1)}{nT}f(\rho_{A})
+2p(1-p)c\frac{1}{T^2}\sum_{i=1}^{T-1}(i+1)^{-\gamma}(T-i). 
\]

\subsubsection{2.2.1) $\gamma>1$ case}
 We can obtain
 \begin{eqnarray}
    V(Z(T)) 
    &\simeq &\frac{p(1-p)}{nT}+
    \frac{p(1-p)(n-1)f(\rho_A)}{nT}
    +\frac{2p(1-p)c}{T^2}\sum_{i=1}^{T-1} (T-i)/(i+1)^{\gamma}\nonumber \\
    &\simeq & 
    \frac{p(1-p)}{nT}+
    \frac{p(1-p)(n-1)\rho_D}{nT}
    +2pqcT^{-\gamma}/(\gamma-1).
\end{eqnarray} 
 The first two terms decrease as $1/T$
 and the third term decreases as  $1/T^{\gamma}$ where $\gamma>1$.
 Hence, the significant terms are the first two terms and the convergence
 of $V(Z(T))$ behaves as  $\sim1/T$.
The convergence speed is the same as that of the independent binomial
case.

\subsubsection{2.2.2) $\gamma=1$ case}
$V(Z(T))$ behaves as
 \begin{equation}
   V(Z(T))\simeq  \frac{p(1-p)}{nT}+
    \frac{p(1-p)(n-1)f(\rho_A)}{nT}
    +\frac{2p(1-p)c}{T^2}\sum_{i=1}^{T-1} (T-i)/(i+1).
    \label{14}
\end{equation} 
The RHS  of Eq.(\ref{14}) is evaluated as 
\begin{equation}
  RHS
  \simeq    
    \frac{p(1-p)}{nT}+
    \frac{p(1-p)(n-1)f(\rho_A)}{nT}
    +2p(1-p)c[(T+1)\log T-T+2]\sim\log T/T.
\end{equation}
 In conclusion,  $V(Z(T))$ behaves asymptotically as
 \begin{equation}
V(Z(T)) \sim \log T/T
 \end{equation}
 and the estimator $Z(T)$ converges to $p$
 more slowly than in the normal case.

 \subsubsection{2.2.3) $\gamma<1$ case}
 $V(Z(T))$ is calculated as:
 \begin{equation}
   V(Z(T))
   \simeq    
    \frac{p(1-p)}{nT}+
    \frac{p(1-p)(n-1)f(\rho_A)}{nT}
    +2p(1-p)c
    [\frac{1}{(1-\gamma)(2-\gamma)T^{\gamma}}]\sim T^{-\gamma}.
\end{equation} 
Then, we can conclude $V(Z(T))$ behaves as
 \begin{equation}
 V(Z(T)) \sim T^{-\gamma}.
\end{equation}

\begin{figure}[htbp]
\begin{tabular}{ccc}    
(a)\includegraphics[width=4.5cm]{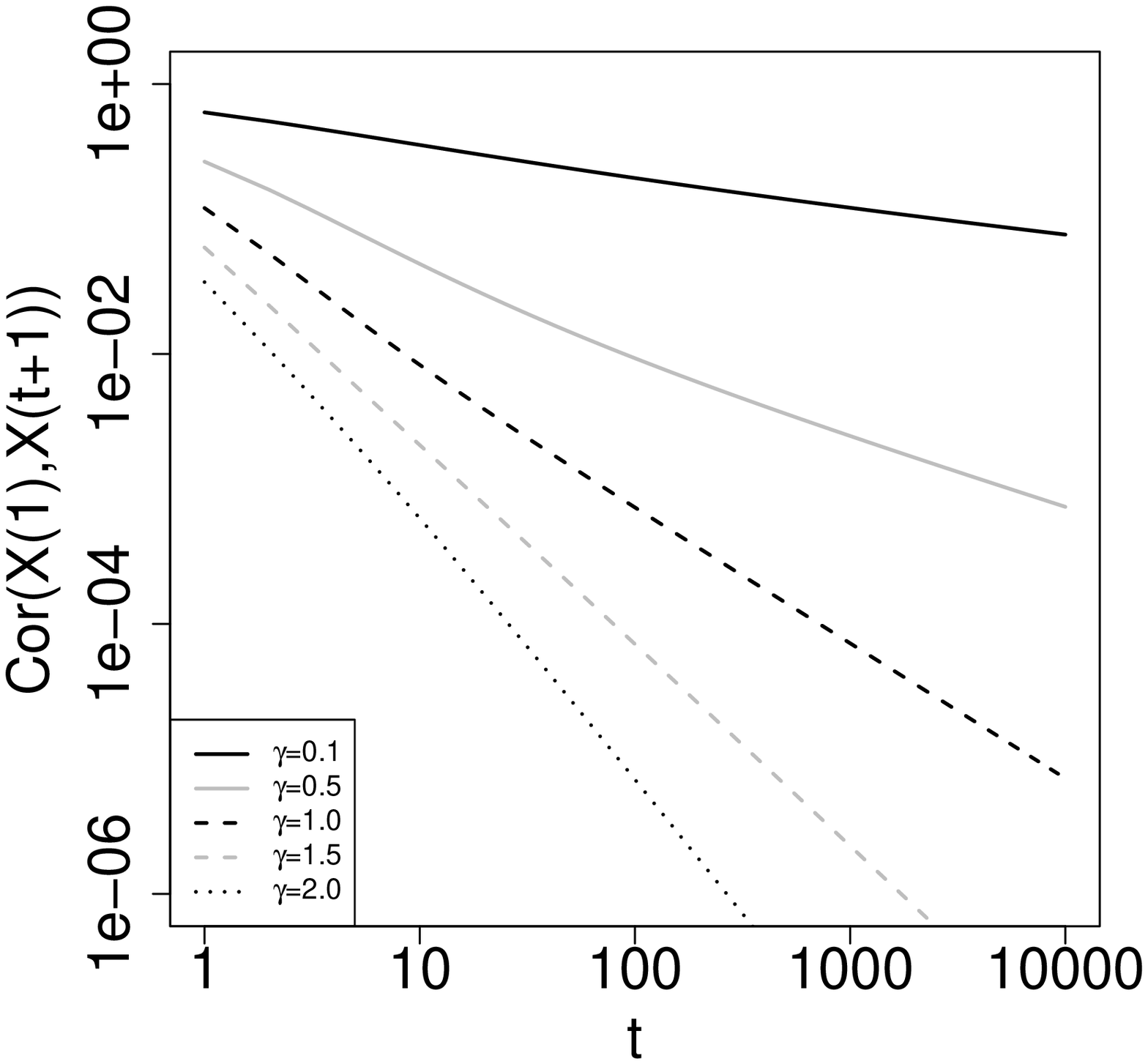} &
(b)\includegraphics[width=4.5cm]{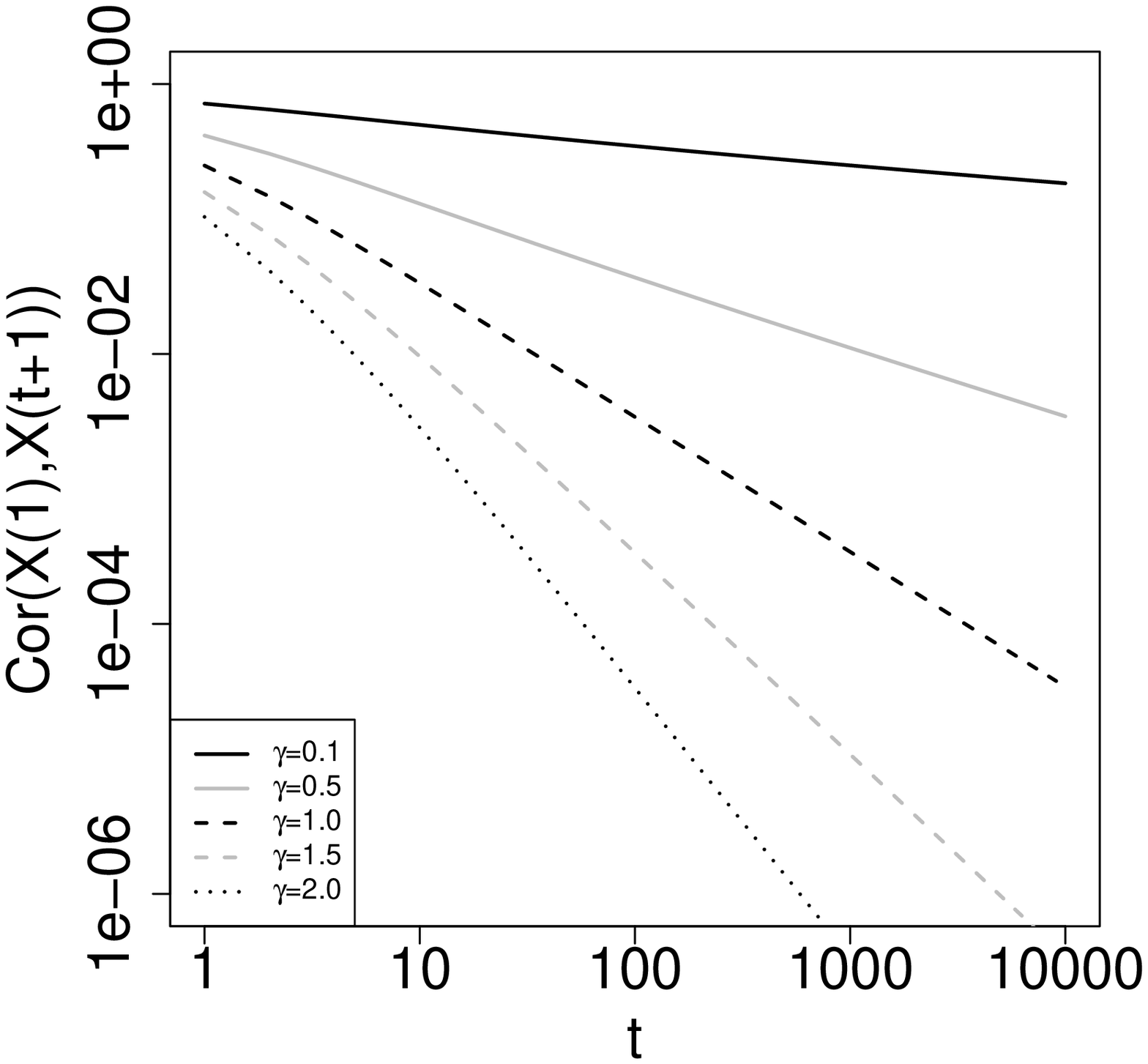} &
(c)\includegraphics[width=4.5cm]{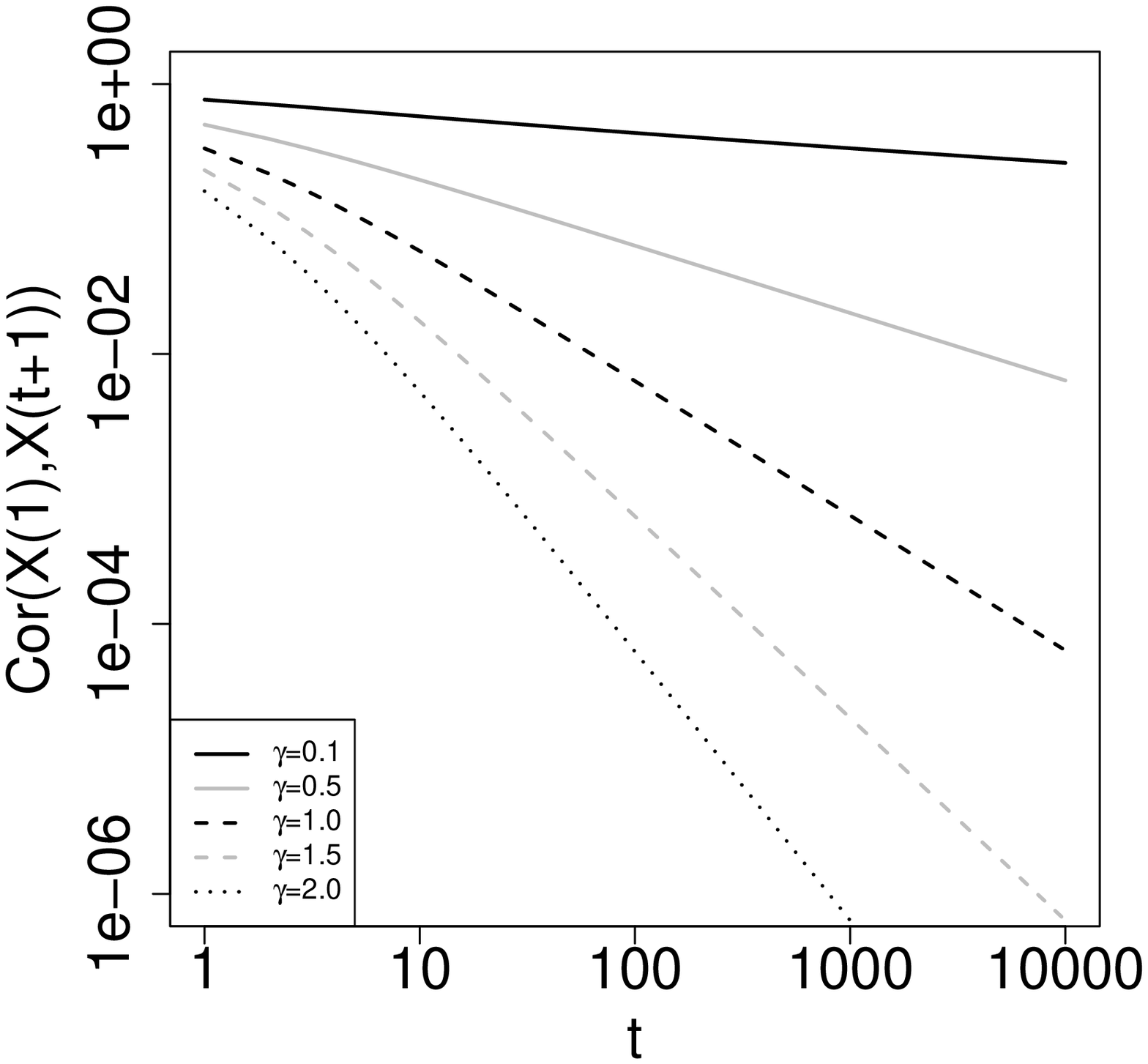} \\
(d)\includegraphics[width=4.5cm]{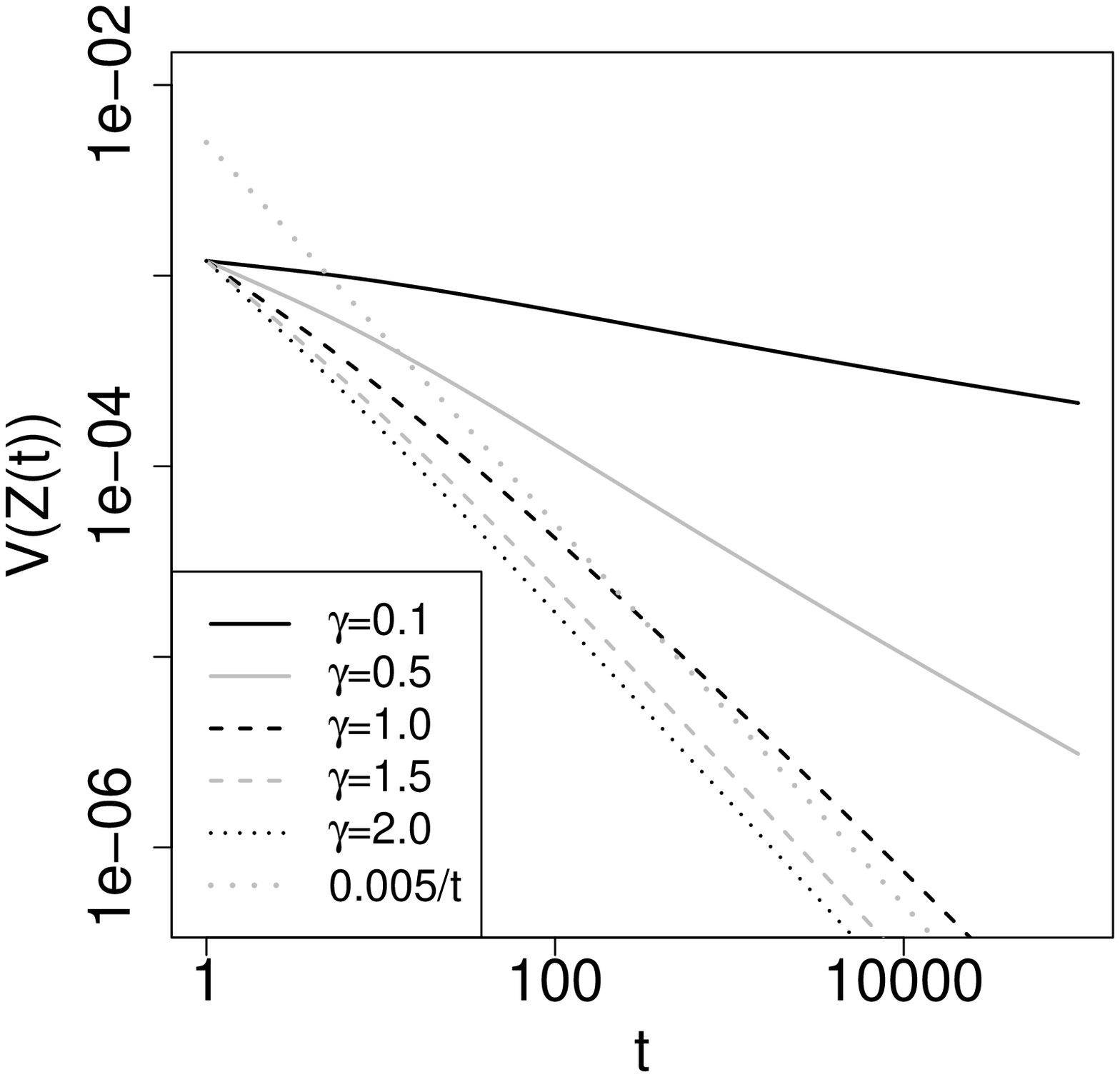} &
(e)\includegraphics[width=4.5cm]{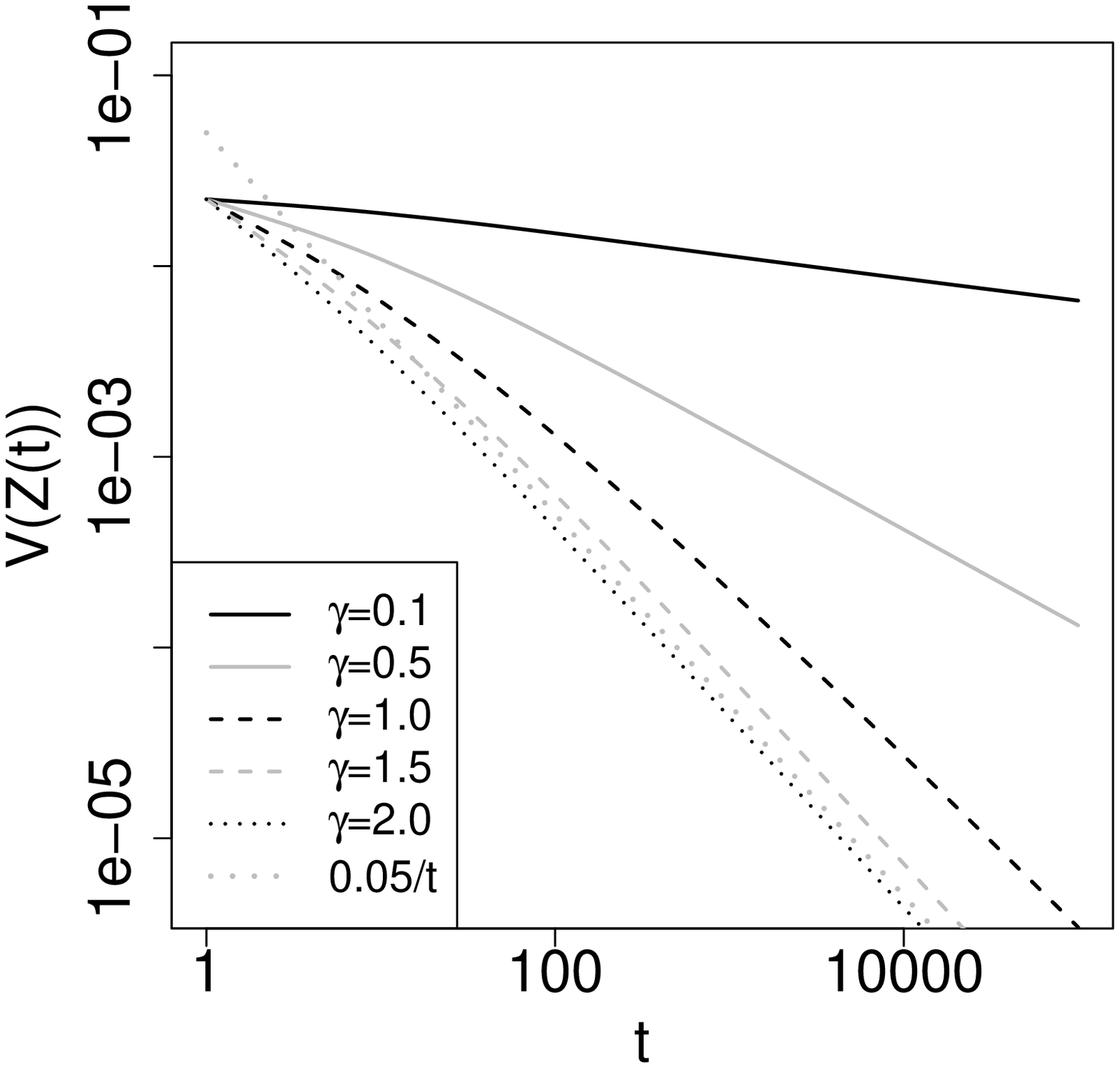} &
(f)\includegraphics[width=4.5cm]{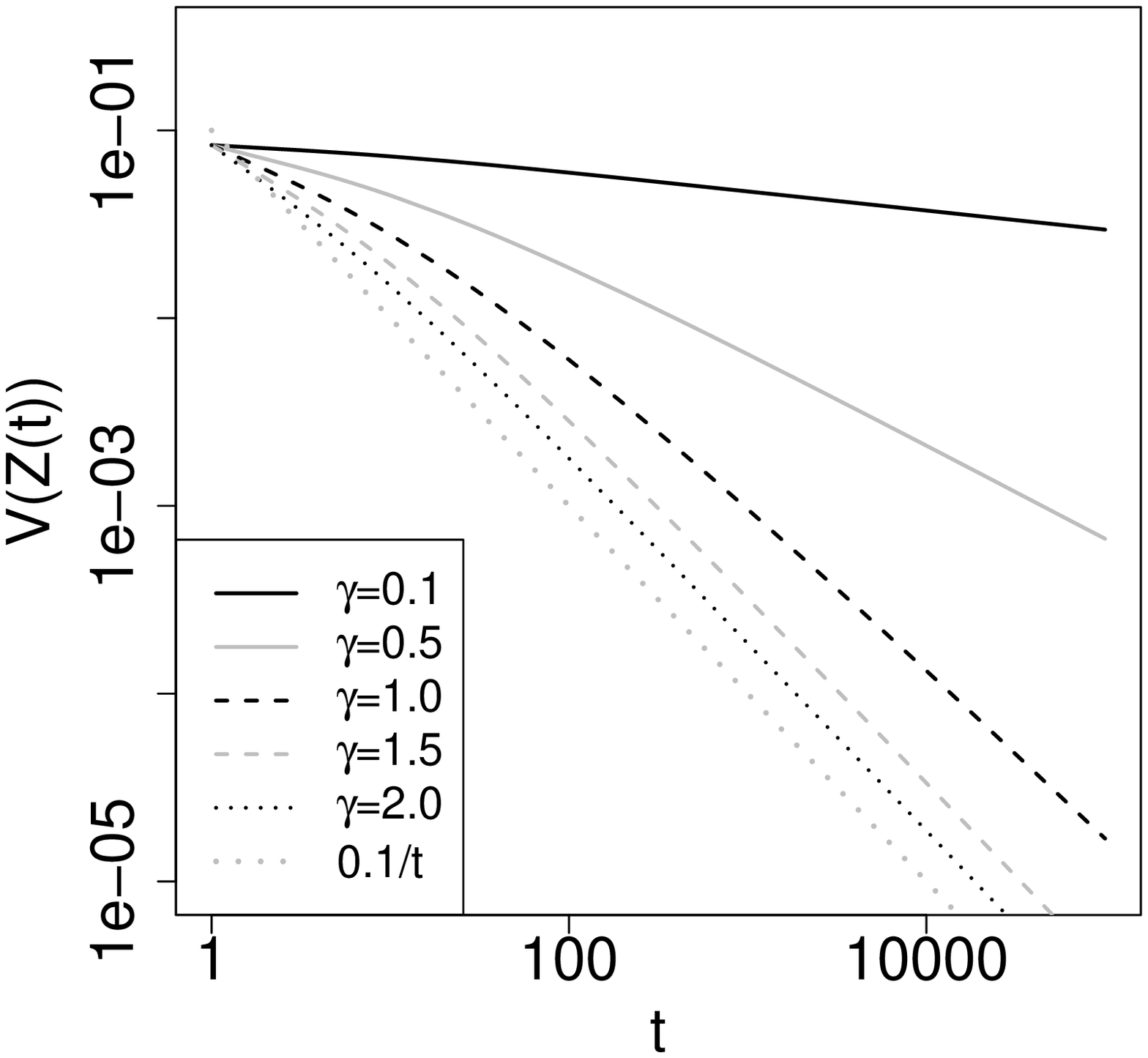} \\
\end{tabular}  
\caption{Plots of (a),(b),(c) $C(t)$ and (d),(e),(f) $V(Z(t))$
  vs. $t$, for
  $\gamma \in \{3.0,2.0,1.5,1.0,0.5,0.1\}$.
  Power decay case. The PDs are  $1\%$ for (a) and (d), $10\%$ for (b) and (e), and $50\%$ for (c) and (f).  
 }
\label{POW}
\end{figure}

In conclusion, a phase transition occurs
when the temporal correlation decays by power law.
When the power index, $\gamma$, is less than one, the PD estimator $Z(T)$
slowly converges to $p$.
Conversely, when the power index $\gamma$ is greater than one,
the convergence behavior is the same as that of the
binomial distribution.
This phase transition is called a "super-normal transition" \cite{Hod,Hisakado2},
which is the transition between long memory and intermediate memory.
This transition is different from the phase transition
found when we used the beta-binomial distribution in our previous work.
In that article, when the power index was less than one, the PD estimator $Z(T)$
did not converge to $p$ when the beta-binomial model was used \cite{Hisakado6}.

\begin{figure}[htbp]
\includegraphics[width=7cm]{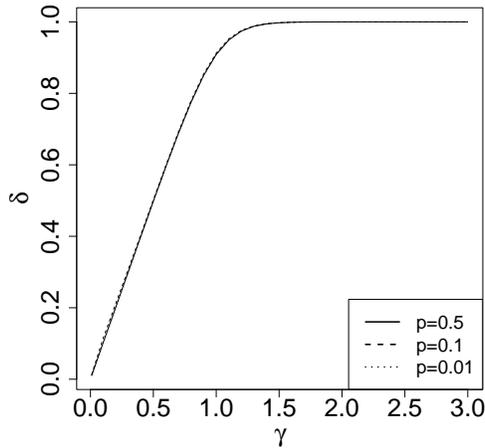} 
\caption{Plots of $\log_{2}(Z(T)/Z(2T))$ with $T=10^5$ vs. $\gamma$.
 The PDs are  $1\%$(dotted), $10\%$(broken) and $50\%$(solid).  
 }
\label{delta}
\end{figure}

To confirm the phase transition, we
calculate $C(t)=f(\rho_{A}(t+1)^{-\gamma})$
and $V(Z(T))$.
Fig. \ref{POW} (a)-(c) shows the double
logarithmic plot of $C(t)$ vs. $t$. $C(t)$ decays by power law for $\gamma \in \{0.1,0.5,1.0,1.5,2.0\}$.
For small $\gamma$, such as $0.5,0.1$, the slope is extremely small.
Fig. \ref{POW} (d)-(f) shows the double logarithmic plot
of $V(Z(t))$ vs. $t$. For $\gamma \in  \{0.1,0.5,1.0,1.5,2.0\}$, $V(Z(t))$ decays as $1/t$.
At $\gamma\leq 1$, the slope of the decay becomes less than one.
In this case, the convergence becomes slower than in the normal case.

Next, we confirm the phase transition using 
  finite size scaling.
 We estimate the exponent of the convergence of $Z(T)$.
If we assume that $V(Z(T))\propto T^{-\delta}$, the exponent $\delta$
is estimated as
\[
\delta=\log_{2}\frac{V(Z(T))}{V(Z(2T))}.
\]
In the case $V(Z(T))\sim \ln T/T$, we have
\[
\log_{2}\frac{V(Z(T))}{V(Z(2T))}=1-\log_{2}(1+\frac{2}{\ln T})<1.
\]
We estimate $\delta$ numerically for $T=10^5$.
We plot the results in Fig. \ref{delta}.
We see that $\delta=1$ for $\gamma>1$ and $\delta=\gamma$ for $\gamma<1$.
When $\gamma\simeq 1$, the relation
becomes obscured by the finite size effect.

In summary, when $\gamma>1$, $Z(T)$ converges to $p$ as in the normal case.
On the other hand, when $\gamma\leq 1$, the convergence is
slower than that of the  normal case.
Hence, there is the phase transition at $\gamma=1$.

\section{3. Estimation of parameters}
As discussed in the previous section, whether 
temporal correlation obeys an exponential decay or a power decay is an
important issue
because there exists a super-normal transition in the latter case.
Further, the appearance of a transition affects whether we can estimate the PD.

\begin{figure}[h]
\begin{center}
\begin{tabular}{c}
% 1
\begin{minipage}{0.5\hsize}
\begin{center}
\includegraphics[clip, width=8cm]{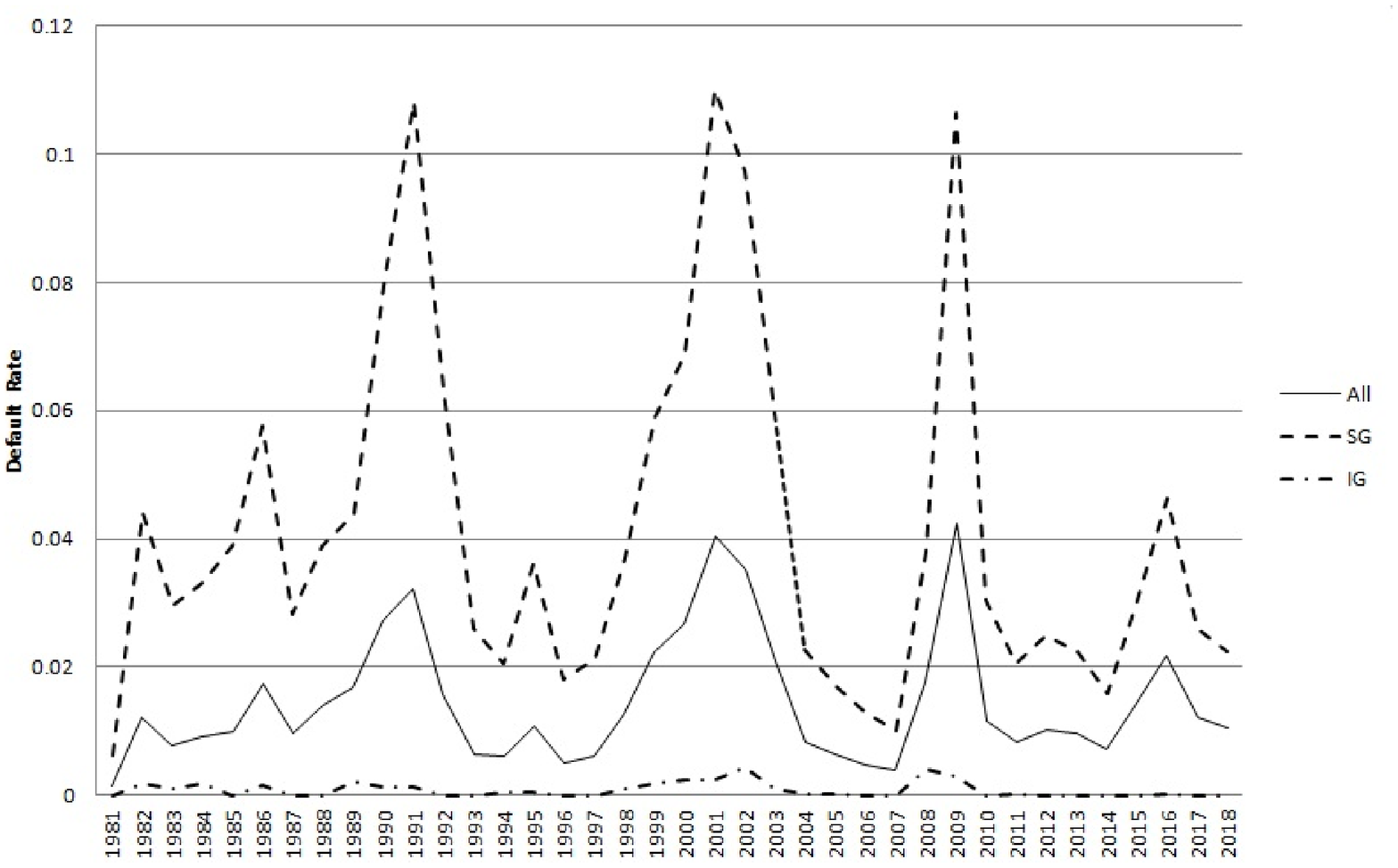}
\hspace{1.6cm} (a)
\end{center}
\end{minipage}
% 2
\begin{minipage}{0.5\hsize}
\begin{center}
\includegraphics[clip, width=8cm]{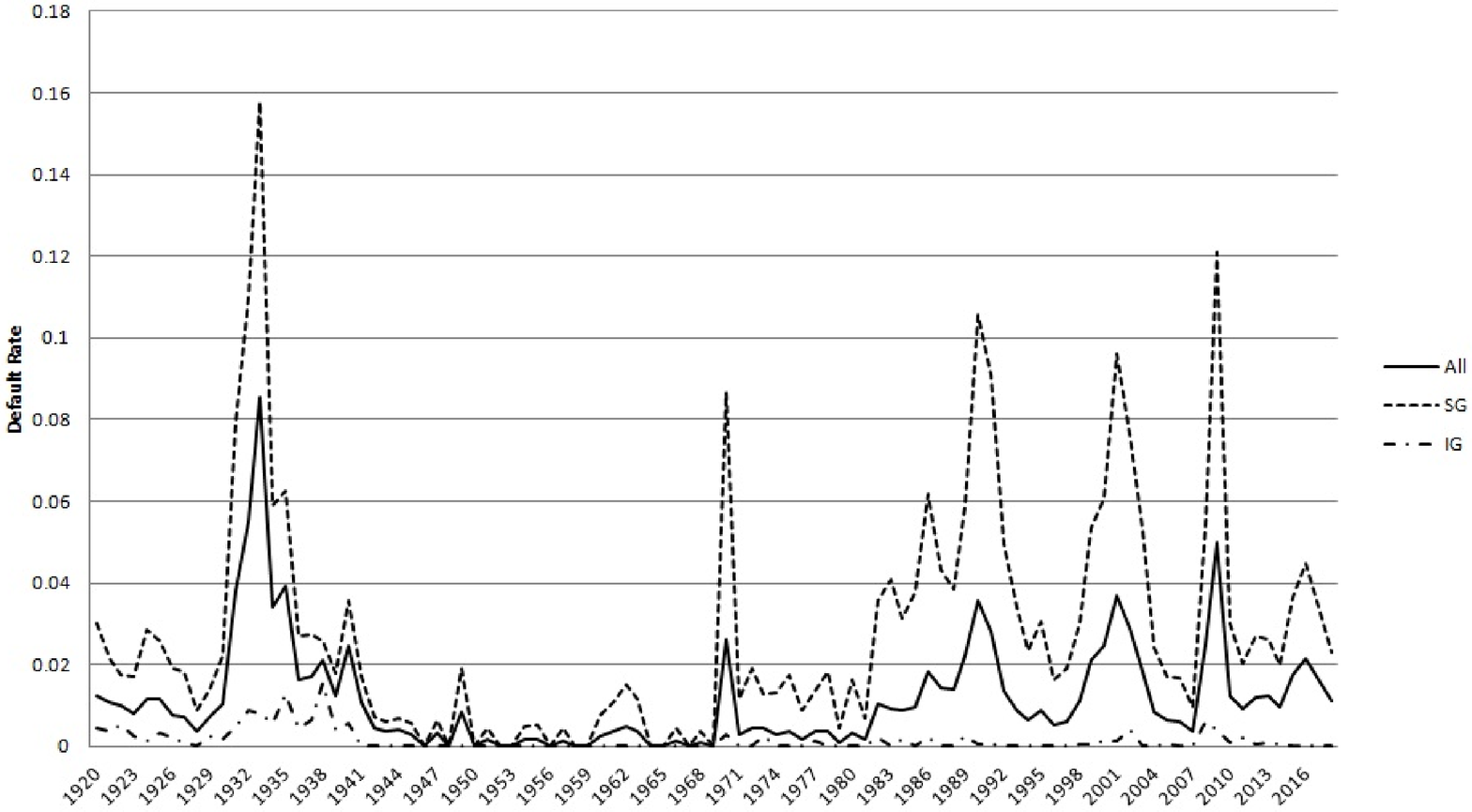}
 \hspace{1.6cm} (b)
\end{center}
\end{minipage}
% 3
 \end{tabular}
\caption{ (a): S\&P Default Rate in 1981-2018. (b)Moody's Default Rate in
  1920-2018.  The solid and dotted lines respectively correspond to 
  the speculative grade (SG) and investment grade (IG) of all the samples. }
\label{3}
\end{center}
\end{figure}

First, the S\&P default data from 1981 to 2018 \cite{Data1} are used. The average PD is 1.51 $\%$ for all ratings,  3.90 $\%$ for speculative  grade (SG) ratings, and 0.09$\%$ for investment grade  (IG) ratings. 
 The SG rating represents ratings under BBB-(Baa3) and IG represents that above BBB-(Baa3).
In Fig. \ref{3} (a) we show the historical default rate of the S\&P.
The solid and dotted lines correspond to the speculative grade and investment grade samples, respectively.
We use Moody's default data from 1920 to 2018 for 99 years \cite{Data2}. It includes the Great Depression in 1929 and Great Recession in 2008.
The average default rate is 1.50$\%$ for all of the ratings, $3.70\%$
 for  speculative ratings, and $0.14\%$ for investment grade.  In Fig. \ref{3} (b), we show the historical default rate of Moody's.

%%%%%%%%% Fig.1

%%%%%%%%%%%%%%%%%% Table3
\begin{table}[tbh]
\caption{MAP estimation of the parameters  for  the exponential and power decay models by the Merton model}
\begin{center}
\begin{tabular}{|l|l|ccc|ccc|r|}
\multicolumn{4}{c}{}\\ \hline
&  & &Exponential decay  &&&Power decay  & \\
No.& Model & 
 $p$& $\rho_D$ &$\theta$& $p$& $\rho_D$ &$\gamma$\\ \hline \hline
1&Moody's 1920-2018& 1.43\%& 17.8\% &0.858& 7.40\%& 38.7\% &0.090\\ \hline
2&S\&P 1981-2018& 1.43\% & 6.4\% & 0.597& 1.85\% & 12.0\% & 0.610 \\ \hline
3& Moody's 1981- 2018& 1.61\% & 7.1\%&0.613& 1.92\% & 12.4\%&0.622 \\ \hline
4&S\&P 1990-2018& 1.72\% & 7.5\% & 0.616& 2.97\% & 12.5\% & 0.616  \\ \hline
5&Moody's 1990-2018& 1.92\% & 10.0\%& 0.678& 2.40\% & 12.1\%& 0.624 \\ \hline
6&Moody's 1920-2018 SG & 3.00\% & 18.9\% & 0.838& 6.15\% & 32.2\% & 0.146  \\ \hline
7&S\&P 1981-2018  SG& 4.53\% & 8.7\% & 0.588& 4.42\% & 11.7\% & 0.628  \\ \hline
8&Moody's 1981-2018 SG&4.28\% & 9.4\% & 0.603 &3.97\% & 11.5\% & 0.619\\ \hline
9&S\&P 1990-2018  SG& 4.93\% & 11.2\% & 0.639& 5.40\% & 13.9\% & 0.626  \\ \hline
10&Moody's 1990-2018  SG& 4.51\% & 11.1\% & 0.648& 6.09\% & 14.6\% & 0.619 \\ \hline
11&Moody's 1920-2018 IG & 0.04\% & 35.3\% & 0.891& 3.40\% & 51.4\% & 0.102  \\ \hline
12&S\&P  1981-2018 IG&0.02\% & 25.8\% & 0.483 &0.02\% & 20.3\% &  9.189\\ \hline
13&Moody's 1981-2018 IG&0.01\% & 21.9\% & 0.672 &1.84\% & 33.8\% &  0.618\\ \hline
14&S\&P 1990-2018  IG& 0.01\% & 37.4\% & 0.712& 1.63\% & 46.7\% & 0.630\\ \hline
15&Moody's 1990-2018  IG& 0.01\% & 33.0\% & 0.794& 3.44\% & 51.1\% & 0.003 \\ \hline
\end{tabular}
\label{game13}
\end{center}
\end{table}

We estimate the parameters $p,\rho_{A},\theta$ and $\gamma$ 
of the Merton model using the Bayesian method and Stan 2.19.2 in R 3.6.2 software.
We explain the method and how to estimate the parameters in Appendix B \cite{Tas2} and
summarize the results in Table \ref{game13}.
We show $\rho_{D}$ instead of $\rho_{A}$, as we need to compare it with that
of the beta-binomial distribution model.
The estimation of the parameters are the maximum a
posteriori (MAP) estimation.
A detailed explanation of the estimation procedure and rmd file is provided on github \cite{git}.  
We notice that the power index $\gamma$ is smaller than 1
for all cases and the values are smaller than the phase transition point, 
$\gamma=1$. 

We compare these results to the MAP estimation using the beta-binomial distribution by using the same data \cite{Hisakado6}.
The conclusions are shown in Table \ref{game3} for the exponential and
 power decay models.
 We confirmed small $\theta$ and large $\gamma$ values, which
represent small temporal correlation.
 The parameter $\gamma$ for the power decay is  larger  than the phase transition point, 
$\gamma=1$. 
 The PD and default correlation are almost the same as the estimations
 by the exponential  and power decay models.
 The reason behind this is that the power exponent $\gamma$ is adequately large
 and there is only a small difference between the exponential and power decay models.

 \begin{table}[tbh]
\caption{Most likelihood estimate of the parameters  for  the exponential and power decay models by beta-binomial distribution }
\begin{center}
\begin{tabular}{|l|l|ccc|ccc|r|}
\multicolumn{4}{c}{}\\ \hline
&  & &Exponential decay  &&&Power decay  & \\
No.& Model & 
 $p$& $\rho_D$ &$\theta$& $p$& $\rho_D$ &$\gamma$\\ \hline \hline
1&Moody's 1920-2018& 0.96\%& 1.9\% &0.044& 0.94\%& 2.0\% &4.7\\ \hline
2& S\&P 1981- 2018& 1.53\% & 0.8\%&0.026& 1.54\% & 0.8\%&5.7 \\ \hline
3&Moody's 1981-2018& 1.53\% & 0.8\% & 0.022& 1.52\% & 0.7\% & 5.9  \\ \hline
4&S\&P 1990-2018& 1.66\% & 0.9\%& 0.023& 1.64\% & 0.9\%& 5.7  \\ \hline
5&Moody's 1990-2018& 1.67\% & 0.9\% & 0.019& 1.61\% & 0.8\% & 6.0  \\ \hline
6&Moody's 1920-2018 SG & 2.36\% & 3.8\% & 0.044& 2.34\% & 4.1\% & 4.7  \\ \hline
7&S\&P 1981-2018 SG&4.16\% & 2.0\% & 0.026 &4.20\% & 2.0\% & 5.7 \\ \hline
8&Moody's 1981-2018  SG& 4.18\% & 2.0\% & 0.022& 4.35\% & 1.9\% & 6.0  \\ \hline
9&S\&P 1990-2018  SG& 4.42\% & 2.5\% & 0.024& 4.43\% & 2.6\% & 5.6  \\ \hline
10&Moody's 1990-2018  SG& 4.33\% & 2.3\% & 0.020& 4.31\% & 2.2\% & 5.9  \\ \hline
11&Moody's 1920-2018 IG & 0.13\% & 0.8\% & 0.17& 0.11\% & 0.9\% & 3.0  \\ \hline
12&S\&P 1981-2018 IG&0.11\% & 0.4\% & 0.12 &0.09\% & 0.3\% & 3.6 \\ \hline
13&Moody's 1981-2018  IG& 0.10\% & 0.6\% & 0.05& 0.09\% & 0.3\% & 4.6  \\ \hline
14&S\&P 1990-2018  IG& 0.09\% & 0.4\% & 0.12& 0.09\% & 0.4\% & 3.7  \\ \hline
15&Moody's 1990-2018  IG& 0.09\% & 0.4\% & 0.06& 0.07\% & 0.7\% &4.2   \\ \hline
\end{tabular}
\label{game3}
\end{center}
\end{table}

We can confirm that $\theta$ and $\gamma$ both have large differences between the values estimated by the beta-binomial distribution and the Merton model.
The reason for this is shown in Fig. \ref{rho} (a).
We set  $d_{1A}$ and $d_{1D}$ for $\rho_A$ and $\rho_D$, respectively.
From this, we can obtain the inequality 
\[
d_{1A}=\frac{d_{1A}\rho_A}{\rho_A}
>>\frac{f(d_{1A}\rho_A)}{f(\rho_{A})}=\frac{f(d_{1A}\rho_A)}{\rho_D}=d_{1D}.
\]
Hence, the difference in the estimated parameter between the  Merton model and the beta-binomial model becomes  large.
We can find the large convexity for the  mapping function $f$.
Hence, $\theta$ and $\gamma$ for a default correlation is much smaller
than  that for asset correlation.

Next, we discuss whether the correlation has a long memory.
In Table \ref{game14}, we  calculated the WAIC and WBIC for each model that uses the Merton model for the discussion.
Using Moody's  data from 1920, the power decay model
is found to be superior to the exponential decay model.
Therefore, it seems that the default rate has a long memory.
As  $\gamma$ is less than 1 for long history data, the phase is
in the slow convergence phase.
In other words, parameter estimation becomes difficult
because the convergence speed becomes slow when the
temporal correlation is the power decay.

In Table \ref{game24}, we show the AIC and BIC for each model using the beta-binomial distribution and compare them to the estimation using Merton model.
We obtain the same conclusion using Moody's data from 1920: the power decay model is superior to the exponential decay model.
The parameter $\gamma$ is not less than 1 for
power decay case when we use the beta-binomial distribution.

\begin{table}[tbh]
\caption{WAIC and WBIC for  the exponential and power decay using the Merton model}
\begin{center}
\begin{tabular}{|l|l|cc|cc|r|}
\multicolumn{4}{c}{}\\ \hline
&  & Exponential decay&  & Power decay &  \\
No.& Model & 
 WAIC& WBIC & WAIC& WBIC \\
 \hline \hline
1&Moody's 1920-2018& 572.9& 746.7& 568.6& 745.9 \\ \hline
2& S\&P  1981- 2018& 271.5 & 332.9&  272.1& 334.8 \\ \hline
3& Moody's 1981-2018& 277.6 & 339.0 & 277.5 & 341.4 \\ \hline
4&S\& P 1990-2018& 214.3 & 256.4& 214.6 & 258.3 \\ \hline
5&Moody's 1990-2018& 219.5 & 262.1 & 219.6 & 264.7 \\ \hline
6&Moody's 1920-2018 SG & 564.3 & 731.9 & 560.2 & 733.8 \\ \hline
7&S\&P 1981-2018 SG&268.7 & 328.1&268.9 & 330.5 \\ \hline
8&Moody's 1981-2018  SG& 274.1& 333.6& 274.4& 337.1  \\ \hline
9&S\& P 1990-2018  SG& 212.0 & 253.9 & 212.8 & 255.2 \\ \hline
10&Moody's 1990-2018  SG& 217.3 & 260.4 & 218.3 & 262.8 \\ \hline
11&Moody's 1920-2018 IG & 247.6 & 351.2& 244.9& 351.2   \\ \hline
12&S\&P 1981-2018 IG&116.0 & 153.3 &115.0 & 160.3 \\ \hline
13&Moody's  1981-2018 IG&110.3& 156.3&108.8 & 156.7\\ \hline
14&S\&P 1990-2018  IG& 87.6 & 114.3 & 86.8 & 117.7   \\ \hline
15&Moody's 1990-2018  IG& 81.7 & 111.4 & 82.5& 114.5  \\ \hline
\end{tabular}
\label{game14}
\end{center}
\end{table}

\begin{table}[tbh]
\caption{AIC and BIC  for  the exponential and power decay using the beta-binomial distribution}
\begin{center}
\begin{tabular}{|l|l|cc|cc|r|}
\multicolumn{4}{c}{}\\ \hline
&  & Exponential decay&  & Power decay &  \\
No.& Model & 
 AIC& BIC & AIC& BIC \\
 \hline \hline
1&Moody's 1920-2018& 746.8& 780.8& 746.8& 780.9 \\ \hline
2& S\&P  1981- 2018& 352.0 & 382.9&  353.6& 384.5 \\ \hline
3& Moody's 1981-2018& 362.2 & 395.0 & 363.4 & 396.2 \\ \hline
4&S\& P 1990-2018& 285.0 & 315.5& 285.8 & 316.3 \\ \hline
5&Moody's 1990-2018& 293.8 & 326.3 & 298.6 & 331.1 \\ \hline
6&Moody's 1920-2018 SG & 730.8 & 762.0 & 730.6 & 761.8 \\ \hline
7&S\&P 1981-2018 SG& 346.8 & 366.7&348.8& 368.7 \\ \hline
8&Moody's 1981-2018  SG& 356.0& 385.9& 360.4& 390.3  \\ \hline
9&S\& P 1990-2018  SG& 283.0 & 302.6 & 283.8 & 303.4 \\ \hline
10&Moody's 1990-2018  SG& 291.0 & 320.7 & 291.6 & 321.3 \\ \hline
11&Moody's 1920-2018 IG & 302.2 & 334.9& 300.4& 333.1 \\ \hline
12&S\&P 1981-2018 IG&134.9 & 164.5 &136.1 & 165.7 \\ \hline
13&Moody's  1981-2018 IG&142.3& 173.7& 140.2 & 171.7\\ \hline
14&S\&P 1990-2018  IG&  106.0 & 135.3 & 107.2& 136.5  \\ \hline
15&Moody's 1990-2018  IG& 111.8 & 142.8 & 112.4& 143.4  \\ \hline
\end{tabular}
\label{game24}
\end{center}
\end{table}

\section{4. Concluding Remarks}

In this study, we introduced  the  Merton model with temporal asset correlation
and discussed the convergence of the estimator of the probability of default.
We adopted a Bayesian estimation method to estimate the model’s parameters and
discussed its implication in the estimation of PD.
We found a phase transition when the temporal correlation decayed by a
power curve, which meant that the correlation had a long memory.
When the power index $\gamma$ was larger  than one, the
estimator distribution of the  PD converged normally.
When the power index was less than or equal to 1, the distribution
converged slowly. This phase transition is called the
"super-normal transition".
For the case of an exponential decay, there was no phase transition.

In our previous work, we studied a beta-binomial distribution model with
temporal default correlation. The estimator of PD also showed a phase transition
in the power  decay case. 
The transition depended on whether the distribution converged or not.
It was different from the phase transition found in the present study.

The main difference between the Merton model and the
beta-binomial distribution model is the incorporation
of default correlation.	
In the latter model, the default correlation is defined by binary variables.
In the former, the default correlation is incorporated into 
the asset correlation, which is defined by a continuous variable.
The implication of the present study is about the difficulty
in the estimation of PD when we adopt the former models.
The estimated power index is in the slow convergence region
of PD. Even with empirical data over a long period of time,
PD is difficult to estimate when we adopt the proposed model.

\appendix

\section{Appendix A. Temporal correlation for the Merton Model}

We consider the  assets of two obligors, $\hat{U}_{1t}$ and $\hat{U}_{2t}$, in year $t$ to confirm temporal correlation.
The assets of  two obligors have the correlation $\rho_A$.
$S_t$ is the global  economic factor that affects the two  obligors at $t$:
\begin{eqnarray}
\hat{U}_{1t}&=&\sqrt{\rho_A} S_t+\sqrt{1-\rho_A}\epsilon_{1 t},
\nonumber \\
\hat{U}_{2t}&=&\sqrt{\rho_A} S_t+\sqrt{1-\rho_A}\epsilon_{2 t},
\end{eqnarray}
where
$\epsilon_1$ and $\epsilon_2$ are the individual factors for the obligors.
Here, there is no correlation among  $\epsilon_1$, $\epsilon_2$, and $S_t$ because they are independent of each other.
In the following year, $t+1$, the assets of the two obligors are $\hat{U}_{1 t+1}$ and $\hat{U}_{2 t+1}$.
The assets have the same correlation, $\rho_A$, through the global factor $S_{t+1}$.   
We can write this as:
\begin{eqnarray}
\hat{U}_{1t+1}&=&\sqrt{\rho_A} S_{t+1}+\sqrt{1-\rho_A}\epsilon_{1t+1},
\nonumber \\
\hat{U}_{2t+1}&=&\sqrt{\rho_A} S_{t+1}+\sqrt{1-\rho_A}\epsilon_{2 t+1}.
\end{eqnarray}
The temporal correlation between $t$ and $t+1$ is $d_1$.
The correlation between $\hat{U}_{1 t}$ and $\hat{U}_{2 t+1}$ is $d_1\rho_A$.
In the same way, we obtain the temporal correlation matrix, Eq.(\ref{matrix}).
It is same as that from the Bayesian estimation, which was introduced in \cite{Hisakado6}, without differentiating between the asset and default correlations.

\section{Appendix B. Bayesian estimation using the Merton model}

In this Appendix we explain the estimate of parameters using the Merton model \cite{Tas2}.
There is a prior belief of the possible value on the PD.
The prior belief is updated by observations while using the prior distribution as a weighting function.
Here, we use the prior function, which is a uniform prior distribution.

To calculate the unconditional probability $P(X_1=k_1, \cdots, X_T=k_T)$, we approximate the solution by Monte Carlo simulations and numerical integration.
Here, the number of obligors and defaults in the $t$-th year are $n_t$  and $k_t$, respectively, and they are observable variables.
The likelihood is 
\begin{equation}
P(X_1=k_1, \cdots, X_T=k_T)\sim\sum_{i=1}^{n}
\prod_{t=1}^{T}\frac{n_t!}{k_t! (n_t-k_t)!}
G(S_t^i)^{k_t} (1-G(S_t^i))^{(n_t-k_t)},
\end{equation}
where $G(S_t^i)$ is defined as the   probability that  an obligor will default in year $t$, which is conditional to the　$i$-th
path realization of all global factors such that
\begin{equation}
 G(S_t^i)=\Phi(\frac{\Phi^{-1}(p)-S_t^i\sqrt{\rho_A}}{\sqrt{1-\rho_A}}),   
\end{equation}
where $\rho_A$ is the asset correlation among obligors within a one year window and $\Phi$ is the cumulative normal distribution.
$S_t^i$ is the  correlated multi-dimensional normal distribution and we use the MAP estimation to estimate the parameters.

We have estimated parameters using a beta-binomial distribution provided in Section 3 and  \cite{Hisakado6}.
One of the differences between using the Merton model and beta-binomial distribution is the default correlation and  the  asset correlation.
The default correlation is defined by binary variables.
On the other hand,  the asset correlation is defined by continuous variables.
The other difference is that one can calculate the parameters analytically when using the beta-binomial distribution.
Hence, it is easier to estimate parameters when using the beta-binomial distribution than when using the Merton model.
In fact, we estimate the parameters to be stable in Section 3,
especially for IG samples, which have small PD.
The estimation of IG samples using the Merton model
 is difficult.

\end{document}